\documentclass[12pt]{article}%
\usepackage{ amsmath, graphics, rotating}%

\begin{document}
\large
\begin{center} {\bf Tensor Potential Description of Matter and Space, III.
Gravitation.} \end{center}
\begin{center}
{Boris Hikin}
\small

E-mail:  blhikin@gmail.com

Tel: 310-922-4752, or 310-826-0209, USA \end{center}
\vskip 1em
\vskip 0.5em
\small

A non-geometrical (but with curved space) theory of gravitation characterized by a vector field representing gravitational
matter and a metric tensor representing space is presented. It is derived from a more general theory of matter and 
space in which matter, described by a 3-index tensor potential, is the prime entity and space is deducible and cannot exist
without matter. Philosophically, instead of geometrization of physics, the new theory advances an idea of 
materialization of space.
The equations that govern the theory of gravitation are deduced from the equations of main theory and have a 
completely different form as compared to Einstein's equations of General Relativity. An exact solution for  
spherical symmetry of a point mass with a Schwarzschild type metric tensor is obtained. 
The subject of gravitational waves and cosmology is discussed.
It is shown that gravitational waves can exist within linear approximation. In this case, the metrics tensor varies only in 
time domain while remaining Euclideanly flat in spatial components.

\baselineskip 1pc
\large
\vskip 3em
{\underline {Introduction}}
\vskip 1em
Though the world of physics will in a few years celebrate the centennial anniversary of Einstein's General Relativity
(GR), the field of gravitation is still an area of intense research. 

Dark matter \cite{i1} - \cite{i3} and Pioneer anomaly \cite{i4} are two recent phenomena, which 
having some difficulty of being explained by GR.
There are, of course, more fundamental reasons for such research: GR's
description of the macro world is incompatible
with other theories that describe the micro world of elementary particles.

Several alternative theories of gravitation have been proposed. The spectrum of mathematical approaches varies:
 from Logunov's flat space Relativistic Theory
of Gravitation \cite{i5}, that completely denies curved space, to Helh's Metric Affine Gravitation theory \cite{i6}, 
that requires 74 variables (64 $\Gamma$s and 10 metric) to describe gravitation alone; from multidimensional
Kaluza (or Kaluza-Klein) type gravitation \cite{i7} to Einstein-Schrodinger nonsymmetrical metric \cite{i8}; and from
conformal gravitation \cite{i9}-\cite{i10} to fourth order and f(R) Lagrangian gravitation \cite{i11}-\cite{i12}.

A common feature of these theories is a desire to "geometrize" physics. That is to say, all
physical forces should be viewed as a consequence of some form of geometrical behavior. Geometry and physics are merged
with geometry being the primary entity and the physics - deducible.

There are also significant efforts to improve or slightly modify General Relativity to give it more physical
meaning. The theories of Brans-Dicke \cite{i13}, Jacobson - Mattingly \cite{i14}, and Bekenstein's TeVeS \cite{i15}
suggest that gravitation should be characterized, along with the metric field, by additional fields of
gravitational matter. 
The characteristic element of these theories is that they consider the form of Einstein's equation ($R_{ij}-1/2Rg_{ij}=T_{ij}$)
describing the gravitational field to be correct. So the modifications of these theories leave 
Einstein's equations practically unchanged.

And yet, as one looks at the roots of GR, one must acknowledge that GR rests on three postulates:
 a) space is curved; b) a test body moves along geodesic lines; c) metric obeys Einstein's equations. 

Most physicists have no problem accepting the first postulate. Indeed, from a
philosophical point of view, the idea of a flat infinitely large fixed space (be it Newton or Minkowski type) 
is not viable or physically justifiable just as the idea of an infinitely flat Earth. Here is a quote from Synge \cite{i16}, page 209:
"The concept of absolute time-space is quite untenable in physics and if Einstein's theory of gravitation is actually
at fault, then  what we need is a modification of that theory, not a return to Newton," or we might add - to
Minkowski.

The second postulate should be derived from equations that describe the matter and metric. Such attempts have been made
\cite{i17}, but it has not  been convincing enough - and it has not been included in any (or the most famous)
 textbooks on GR \cite{i16},\cite{i18},\cite{i19}. 
The expression "moves along geodesic lines" implies that the position of a test body can be characterized 
by four coordinates. Such a view on the test body is a classical idealization and totally contradictory to the
 existing view on an elementary particle (proton, electron, etc),
which is rather a field. In any case, this postulate can be used equally in any theory with curved space.

The third postulate, Einstein's equations, seemingly the most perfect and elegant element of GR, is in
fact the most difficult part to accept. It provides no expression for the energy-momentum tensor. 
Its most significant part - the law of conservation ($T^j_{i;j}=0$) - leads to no conserved value. Per Berhoff theorem,
it has no spherical
symmetric time dependence (or radiation). It allows the existence of empty space which contradicts our existing
 philosophical notions. 
It is not compatible with other physical theories (e.g. quantum mechanics) and thus cannot be extended 
into the micro world.

Philip D. Mannheim recently \cite{i9} in his "Shortcomings of Einstein Gravity" raised the question of 
the uniqueness of Einstein GR:
" ...we need to ask whether the Einstein theory is in fact the
only theory which then meets the three classic tests. 
Beyond this, we also note that when Einstein gravity is extended beyond its solar system
origins, no matter in which way it is extended additional concerns arise. When Einstein
gravity is extended to galactic distance scales we get the dark matter problem. When
Einstein gravity is extended to cosmological distance scales we get the cosmological constant
or dark energy problem. When Einstein gravity is extended to strong gravitational fields we
get the singularity problem. And finally, when Einstein gravity is extended far off the mass
shell we get the renormalization problem. For none of these problems is there as yet any
solution which has been experimentally validated. The case for dark matter is made solely
by assuming the a priori validity of Einstein gravity and then arguing that its failure to fit
astrophysical data with luminous sources alone is evidence for the existence of dark matter,
dark matter which has yet to be directly detected in any of the extensive dark matter
searches which have been going on for many years now. The cosmological constant problem
is even more severe since the needed value as inferred from the application of standard
gravity to cosmology is 60 to 120 orders of magnitude less than the value suggested by
microscopic elementary particle physics. With regard to singularities, not only are there no
data which provide direct evidence for the existence in nature of event horizons or trapped
surfaces (or even whether the mass concentrations in galactic centers have radii less than
their Schwarzschild radii), it is not clear whether the existence of singularities in the fabric
of spacetime is a property of nature or an indication of the breakdown of the theory. Finally,
to resolve the renormalizability problem it has been found necessary to generalize the theory
to a superstring theory which introduces two further ingredients for which there is also no
experimental evidence, namely the existence of ten spacetime dimensions and the existence
of a supersymmetry which gives all known particles as yet undetected superpartners."

It is these difficulties of GR that fuel the attempts of many physicists to look for alternative theories and alternative
points of view with regard to space, matter and gravitation.

\vskip 2em
It is in this spirit we proposed \cite{i20} a non geometrical (but with curved space) theory 
 in which the matter is the prime entity and the space is deducible and cannot exist without the matter.
Philosophically, instead of geometrization of physics, the new theory advances an idea of materialization of space.

Based on that theory, in this paper we consider a theory of gravitation,
in which gravitational field is described by two entities: metric tensor $g_{ij}$ and a vector field $G_i$
with the condition $G_iG_jg^{ij}=1$.
Such a representation of gravitational field does not seems to be new \cite{i14}, but is derived from a completely
different view on space and is governed by a completely different set of equations. In fact the condition $G_iG_jg^{ij}=1$, 
as we will see later, is deducible from the main theory as opposed to postulating it ad hoc as it is done in \cite{i14}.

Without repeating the context of the first paper \cite{i20}, let us present here the main points of that theory.

\newpage
\vskip 3em
{\underline {Main Theory}}
\vskip 1em

Matter is described by a 3-index tensor $P^i_{jk}$, which is a principle variable of this theory.
The space is formed (created) by the existence of the matter and is defined by the same variables  $P^i_{jk}$
 that describe the matter. There is only one way by which the metric tensor of the space ($g_{ij}$)
can be defined through $P^i_{jk}$ - it is by the quadratic expression:
\begin{eqnarray}
\label{f1}
&&g_{ij}:=K_{ij}(P^i_{jk})=\nonumber\\
&&\quad\quad\,\,\,\gamma_1 {{\bar P}^m}_{ni}{{\bar P}^n}_{mj}+\gamma_2\bar P_m{{\bar P}^m}_{ij}+
\gamma_3\bar P_i\bar P_j+\nonumber\\
&&\quad\quad\,\,\,\gamma_4 {{\hat P}^m}_{ni}{{\hat P}^n}_{mj}+\gamma_5(\hat P_i \bar P_j+\hat P_j \bar P_i)+
\gamma_6\hat P_i\hat P_j+\nonumber\\
&&\quad\quad\,\,\,\gamma_7 ({{\bar P}^m}_{ni}{{\hat P}^n}_{mj}+{{\bar P}^m}_{nj}{{\hat P}^n}_{mi})+\gamma_8\hat P_m{{\bar P}^m}_{ij}
\end{eqnarray}
where $\bar P^i_{jk}$ and $\hat P^i_{jk}$ are the symmetrical and the anti-symmetrical in low indices
 parts of tensor $P^i_{jk}$ and 
$\bar P_m$, $\hat P_m$ are contractions of the corresponding tensors. The coefficients $\gamma_1$ through $\gamma_8$
are constants to be identified later. The main requirement of this theory is that $g_{ij}$ must have an inverse,
so that $g_{ij}$ could represent a certain space.

This definition immediately produces the following results: 1) space exists only where there is matter ($P^i_{jk}$); 
2) nowhere matter ($P^i_{jk}$) can be equal to zero (otherwise space vanishes).  In general, space is curved (since the 
tensor $P^i_{jk}$ is a function of coordinates).  However, if the $P^i_{jk}$ is constant in some area, the space
of that area is flat. Similarly, if $P^i_{jk}$ changes slowly in some area (say $P^i_{jk}$ is defined by our galaxy 
within our solar system), then the space is almost flat. 

Having defined $g_{ij}$, we can calculate the covariant derivative $P^i_{jk;l}$, and construct a Lagrangian
that is a function of $P^i_{jk;l}$,  $P^i_{jk}$ and $g_{ij}$, which of course in the end is a function of $P^i_{jk}$ only.
The equations of motion for independent variables $P^i_{jk}$ are obtained through variation of the Lagrangian.

One can make $g_{ij}$ an independent variable by adding to Lagrangian the constraints of eq. (\ref{f1}) and 
Lagrange coefficients $T^{ij}$. 
\begin{equation}
\label{f2}
S=\int\nolimits \{L_M(P^i_{jk}, P^i_{jk;l}, g_{ij})+T^{ij}[g_{ij}-K_{ij}(P^i_{jk})]{\,} \}{\sqrt{g}} d^4x
\end{equation}
Everywhere below we will refer to function $L_M$ as "matter Lagrangian" and to the term proportional to $T^{ij}$ -
denoted $L_T$ - as "constraints". The physical meaning of tensor $T^{ij}$ is - up to a sign - the energy-momentum tensor.
This comes directly from the equation obtained by variation of an action integral with respect to $g_{ij}$:
\begin{equation}
\label{f3}
\frac{1}{\sqrt{g}}\delta (\sqrt{g}L_M)/\delta g_{ij}+T^{ij}=0
\end{equation}

Identification of specific fields, such as gravitational field or electromagnetic field is based
on the symmetry of tensor potential $P_{ijk}$, obtained from $P^i_{jk}$ by lowering
its first index.  If we assume that $P_{ijk}$ has no a priori symmetry, then $P_{ijk}$
can be decomposed on seven sub-fields as shown below (for details see \cite{i20}):
\begin{eqnarray}
\label{f4}
&&P_{ijk}=[S_{ijk}+1/6(G_ig_{jk}+G_kg_{ij}+G_jg_{ki})] + \nonumber\\
&&+[(C_{jki}+C_{kji})+1/3(D_jg_{ik}+D_kg_{ij}-2D_ig_{jk})]\nonumber\\
&&+ [H_{ijk}+1/3(F_kg_{ij}-F_jg_{ik}) ]+ 1/6{\epsilon}_{ijkm}E_m 
\end{eqnarray}
where $\epsilon_{ijkl}$ is a fully anti-symmetrical tensor of Levi-Civita.

Out of seven sub-fields, four are vectors ($G_i$, $D_i$, $F_i$ and $E_i$), 
one is a traceless fully symmetrical 3-index tensor $S_{ijk},\,\,S^i_{ik}=0$ and 
two are fully traceless torsion type tensors $C_{ijk}$ and $H_{ijk}$
\begin{eqnarray}
\label{f3a}
&&C_{ijk}=-C_{ikj};\,\, C_{ijk}g^{ij}=0;\,\, C_{ijk}\epsilon^{ijkl}=0\nonumber\\
&&H_{ijk}=-H_{ikj};\,\, H_{ijk}g^{ij}=0;\,\, H_{ijk}\epsilon^{ijkl}=0\nonumber
\end{eqnarray}

Not every sub-field can exist by itself, because not every sub-field can form the
metric tensor $g_{ij}$. In particular, if we consider vector fields, two of them, $F_i$ and $E_i$, cannot form the space
by themselves alone.
For example, if $E_i$ exists by itself then $P_{ijk}=\epsilon _{ijkl}E^l$. Substituting this in eq. (1) we get:
\begin{eqnarray}
\label{f5}
&&g_{kl}=\gamma_8{\epsilon}_{ijkm}E^m{\epsilon}^{ijsn}g_{sl}E_n={\gamma}_8(g_{kl}-E_kE_l)\nonumber\\
&&or\quad g_{kl}(1-\gamma_8)=E_kE_l
\end{eqnarray}
From this it follows that $det(g_{kl})=0$, and the tensor $g_{kl}$ does not have an inverse - which violates 
the main requirement.

The other two vector fields ($G_i$ and $D_i$), in general, can form a space. But for that to occur $\gamma$s must satisfy 
certain relations and there is no set of $\gamma$s that both vector fields form the space simultaneously - 
it is only one or the other.

In this paper (as in the first one \cite{i20}) we choose the vector field $G_i$ to be 
identified as the gravitational field. If we say that matter in some area consists
only of gravitational field, then we state that the tensor potential $P^i_{jk}$ is expressed through
vector field $G_i$ and metric tensor $g_{ij}$ by this formula:
\begin{equation}
\label{f5a}
P_{ijk}=P^m_{jk}g_{mi}=\frac16 (G_ig_{kj}+G_jg_{ik}+G_kg_{ij})
\end{equation}
Substituting eq. (\ref{f5a}) in (\ref{f1}) we get:
\begin{equation}
\label{f6}
g_{ij}=(\frac{2\gamma_1}{36}+\frac{\gamma_2}{6})G_kG^kg_{ij}+\frac{10\gamma_1+12\gamma_2+36\gamma_3}{36}G_iG_j
\end{equation}

In order for eq. (\ref{f6}) above to be consistent with the requirement for the metric tensor to represent space,
the $\gamma$s must satisfy these requirements:
\begin{eqnarray}
\label{f7}
&&10\gamma_1+12\gamma_2+36\gamma_3=0\quad
\end{eqnarray}
and
\begin{eqnarray}
\label{f7a}
&&G_iG^i(\frac{2\gamma_1}{36}+\frac{\gamma_2}{6})=1
\end{eqnarray}
From eq. (\ref{f7a}) follows that $G_iG^i$ as a function of coordinates is a constant, which we chose to be 1.
This will make eq. (\ref{f7a}) to have this form: 
\begin{eqnarray}
\label{f7b}
&&\frac{2\gamma_1}{36}+\frac{\gamma_2}{6}=1 \quad or \quad \frac{1}{6}(\frac{\gamma_1}{3}+\gamma_2)=1
\end{eqnarray}

The choice of the constant for $G^iG_i=const$ to be one is a matter of convenience and could be 
replaced by any other constant, resulting in scaling $\gamma_1,\, \gamma_2\,\gamma_3$ by that constant.

It is worth pointing out that the constraint on gravitational vector $G_i$ ($G^iG_i=1$) is not a separate
postulate, but is strictly derived from the theory.

To these two conditions we will later (see section "Gravitation") add one more: 
\begin{eqnarray}
\label{f8}
&&\frac{2\gamma_1}{3}+\gamma_2=0
\end{eqnarray}
which makes the choice of $\gamma$s fully defined for the theory of gravitation.
Solving eq.(\ref{f7}, \ref{f8}) we get: $\gamma_1=-18$, $\gamma_2=12$, $\gamma_3=1$ leaving the
other $\gamma$s undefined. In \cite{i20} we used a requirement of uniqueness of representation (\ref{f5a}), 
which produced the same $\gamma$s. 

The equations of motion that define $G_i$ and metric $g_{ij}$ are derived using the variational principle.
Since the Lagrangian is a function of tensor $P^i_{jk}$, to get the equation of motion for $G_i$,
the variation first must be taken with respect to $P^i_{jk}$ and then $P^i_{jk}$  must be set 
to its expression through $G_i$
per eq. (\ref{f5a}). This will produce 64 equations for only 14 unknown functions - four $G_i$ and ten $g_{ij}$.
In order for the equations to have a non trivial solution the Lagrangian must have some special form.

In general, if we write the Lagrangian $L_M$ as a function of sub-fields ($G_i$, $S_{ijk}$, etc.), it could
be split into two groups: a) "independent sub-fields" - when the Lagrangian depends on one sub-field only, 
and b) "interactions of sub-fields" - when the Lagrangian contains two or more sub-fields.
For example, if we take Lagrangian to be $L_M=\bar P^s_{sk;l}\bar P^t_{tp;q}g^{kp}g^{lq}$, then in terms of sub-fields 
per eq. (\ref{f4}) we will get:
\begin{equation}
\label{f9}
L_M=G^{k;l}G_{k;l}+D^{k;l}D_{k;l}+2G^{k;l}D_{k;l}
\end{equation}
The first two terms of eq. (\ref{f9}) are "independent sub-fields" and the last term corresponds to the
interaction of the G-field and the D-field.

We will now formulate the major physical postulate that will significantly limit the form of $L_M$ and
assure the compatibility of equations of motion for all sub-fields. We will call this postulate: 
"independence of sub-fields".

\vskip 1em
\underline{Independence of Sub-fields Postulate}:
\vskip 0.5em
The interaction of sub-fields is due to the metric tensor $g_{ij}$ and tensor $T^{ij}$ only.
The tensor $T^{ij}$ by itself is due to a variation of the Lagrangian $L_M$ per $g_{ij}$ - see eq.(\ref{f3}).

Mathematically, this means that the Lagrangian $L_M$ can be written as the sum of seven
Lagrangians, each of which depends only on one of seven sub-fields:
\begin{eqnarray}
\label{f10}
&&L_M=L_G(G_i)+L_D(D_i)+L_F(F_i)+L_E(E_i)+\nonumber\\
&&\quad\quad L_S(S_{ijk})+L_C(C_{ijk})+L_H(H_{ijk})
\end{eqnarray}

There is another way to look at this "independence of sub-fields" postulate.
With this postulate, we require that the equations of motion (${Q_i}^{jk}:=\delta L_M/\delta P^i_{jk}=0$)
have the same symmetry as the tensor potential $P^i_{jk}$. In other words, if we separate $Q^i_{jk}$ in 
sub-fields by formula (\ref{f4}), the sub-fields of $Q^i_{jk}$ are the functions of the corresponding sub-fields of $P^i_{jk}$.
For  example, if we take the full symmetrization of field $Q_{ijk}$ and take its contraction, the obtained vector field
is depended only on sub-field $G_i$.

We can take one step further and assume that all similar sub-fields have identical Lagrangian dependence.
Thus all vector sub-fields ($G_i$, $D_i$, $F_i$ and maybe even $E_i$) have their corresponding Lagrangians
obtained form one generic Lagrangian. The same should be applied to Lagrangians for sub-fields $C_{ijk}$
and $H_{ijk}$, since both of these sub-fields have the same symmetry.

\baselineskip 1pc
\vskip 3em
{\underline {Gravitation}}
\vskip 1em
Our goal here is to show that within a framework of the theory described above, there is a Lagrangian $L_G$ of
gravitational matter that yields a Schwarzschild solution for the metric tensor $g_{ij}$. 
In fact, we will show that there is a class of such Lagrangians. Let us choose a Lagrangian in the form:
\begin{eqnarray}
\label{f11}
&&L=L_G+L_T=\{\lambda_1 G_{m;n}G^{m;n}+\lambda_2G_{m;n}G^{n;m}+\lambda_3({G^m}_{;m})^2\nonumber\\
&&\quad\quad\quad\quad\quad\quad+T^{mn}[g_{mn}-K_{mn}(P^i_{jk})]\}
\end{eqnarray}
with the condition $\lambda_1+\lambda_2+\lambda_3=0$. 
$K_{mn}$ is the quadratic function of $P^i_{jk}$ given by expression (\ref{f1}).
$G_i$ is defined by eq (\ref{f4}) and should be viewed as a function of $P^i_{jk}$,
which contains only symmetrical part $\bar{P^i_{jk}}$: 
\begin{equation}
\label{f12}
G_i=1/3(2\bar {P^i_{jk}}g^j_i+\bar {P^m_{jk}}g_{mi}g^{jk})
\end{equation}

The first set of equations of motion is obtained by varying $L_G$ by $P^i_{jk}$.  For the variation of $L_G(G_i)$ we get:
\begin{eqnarray}
\label{f13}
&&\delta L_G/\delta P^i_{jk}=\delta L_G/\delta G_n\,{^.}\,\delta G_n/\delta P^i_{jk}\nonumber\\
&&=Q^n(1/3g^k_ng^j_i+1/3g^j_ng^k_i+1/3g_{ni}g^{jk})\nonumber\\
&&=1/3(Q^kg^j_i+Q^jg^k_i+Q_ig^{jk})
\end{eqnarray}
Here $g^i_j$ is the Kronecker symbol and $Q^n$ is the variation of $L_G$ with respect to $G_n$.

\begin{eqnarray}
\label{f14}
&&Q^n=\partial L_G/\partial G_n-(\partial L_G/\partial G_{n;s})_{;s}=\nonumber\\
&&-2\lambda_1 {G^{n;s}}_{;s}-2\lambda_2 {G^{s;n}}_{;s}-2\lambda_3 {{G^s}_{;s}}^{;n}
\end{eqnarray}
Let us note here that if we lower all indices in expression (\ref{f13}) we will get a fully symmetrical tensor.
This, of course, is guaranteed by the choice of the Lagrangian as a function of only $G_i$.

Let us now consider the variation of the constraint Lagrangian $L_T$ by $P^i_{jk}$.
For the purpose of gravitation, it is sufficient to assume that $L_T$ depends only on the symmetrical part of the
tensor $P^i_jk$ ($\bar P^i_{jk}$) or that $\gamma_4$ through $\gamma_8$ are all zero.

Variation of the quadratic function (eq. \ref{f1}) is simply a partial derivative of that function by $P^i_{jk}$:
\begin{eqnarray}
\label{f15}
&&{{Q_{(T)}}_i}^{\,jk}:=\partial L_T/\partial P^i_{jk}=-[\gamma_1(T^{km}\bar P^j_{mi}+T^{jm}\bar P^k_{mi})+\nonumber\\
&&+1/2\gamma_2(T^{mn}\bar P^k_{mn}g^j_i+T^{mn}\bar P^j_{mn})g^k_i)+\gamma_2T^{jk}\bar P_i+\nonumber\\
&&+\gamma_3(T^{mk}\bar P_mg^j_i+T^{mj}\bar P_mg^k_i)]
\end{eqnarray}
Substituting for $\bar P^i_{jk}$ its expression through $G_i$, eq (\ref{f5a}), and lowering all indices we get
\begin{eqnarray}
\label{f16}
&&Q_{(T)\,ijk}=-[(\frac{\gamma_1}{6}+\frac{\gamma_2}{6}+\gamma_3)(T_kg_{ij}+T_jg_{ik})+\frac{\gamma_1}{6}(T_{ik}G_j
+T_{ij}G_k)\nonumber\\
&&+(\frac {\gamma_1}{3}+\gamma_2)T_{jk}G_i+\frac{\gamma_2}{12}(TG_jg_{ik}+TG_kg_{ij})]
\end{eqnarray}
Here, $T_k=T_{km}G^m$ and $T=T_{mn}g^{mn}$

It is not difficult to show that in order to be fully symmetrical as in eq (\ref{f13}), $T^{ij}$ should have this form:
\begin{eqnarray}
\label{f17}
&&T^{ij}=T(k_1g^{ij}+k_2G^iG^j)
\end{eqnarray}
where constants $k_1$ and $k_2$ satisfy the condition $4k_1+k_2=1$.
Substituting (\ref{f17}) into (\ref{f16}), we get
\begin{eqnarray}
\label{f18}
&&Q_{(T)\,ijk}=-[Tk_2(\frac {2\gamma_1}{3}+\gamma_2)G_iG_jG_k\nonumber\\
&&+(\frac{\gamma_2}{12}+\frac{\gamma_1 k_1}{6})(T(G_jg_{ik}+G_kg_{ij})+(\frac{\gamma_1}{3}+\gamma_2)k_1G_ig_{jk}]
\end{eqnarray}
In order for expression (\ref{f18}) to be in the form of expression (\ref{f13}), two conditions must be met: 
\begin{eqnarray}
\label{f19}
&&k_2(\frac {2\gamma_1}{3}+\gamma_2)=0\quad and\nonumber\\
&& (k_1+k_2)(\frac{\gamma_1}{12}+\frac{\gamma_2}{12}+\gamma_3)+
\frac{\gamma_2}{12}+\frac{\gamma_1k_1}{6}=(\frac{\gamma_1}{3}+\gamma_2)k_1
\end{eqnarray}
For the first condition to be true, we have two possibilities: $k_2=0$ or $(\frac {2\gamma_1}{3}+\gamma_2)=0$.
In the first possibility, the requirement $k_2=0$ leads to $k_1=\frac14$ and $\gamma_2=2\gamma_3$. This, with
eq. (\ref{f7}), leads to $\gamma_1=-3\gamma_2$ which
contradicts eq. (\ref{f7b}). In the second possibility, the equation
$\frac {2\gamma_1}{3}+\gamma_2=0$, along with two other equations - (\ref{f7}) and (\ref{f7b}) -
produces a full set of equations for three
unknown $\gamma$s, which is satisfied if $\gamma_1=-18$, $\gamma_2=12$ and $\gamma_3=1$. Indeed:
\begin{eqnarray}
\label{f20}
&&10\gamma_1+12\gamma_2+36\gamma_3=0 \quad for \,\, term\,\, G_iG_j \,\, to\,\, vanish\nonumber\\
&&\frac{2\gamma_1}{36}+\frac{\gamma_2}{6}=1 \quad with \,\,condition\,\, G_kG^k=1\nonumber\\
&& \frac {2\gamma_1}{3}+\gamma_2=0
\end{eqnarray}
Using these values for $\gamma s$ we get $k_1=\frac{1}{9}$ and $k_2=\frac{5}{9}$ and the equations for $Q_{ijk}$
have this form:
\begin{eqnarray}
\label{f21}
&&Q_{ijk}= -\frac{2}{3}(Q_ig_{jk}+Q_jg_{ik}+Q_jg_{ij})-\frac{2}{3}T(G_ig_{jk}+G_jg_{ik}+G_jg_{ij})=0\nonumber\\
&& or\quad Q_i+TG_i=0\nonumber\\
&& or\quad \lambda_1 {G_{i;s}}^{;s}+\lambda_2 {G_{s;i}}^{;s}+\lambda_3 {G^s_{;s}}^{;i}+TG_i=0
\end{eqnarray}
We now replace $\lambda_2$ with $-\lambda_1-\lambda_3$ to get the final form for the first set of equations:
\begin{eqnarray}
\label{f22}
&&\lambda_1 ({G_{i;s}}^{;s}-{G_{s;i}}^{;s})+\lambda_3 ({G^s_{;s}}^{;i}-{G_{s;i}}^{;s})+TG_i=0
\end{eqnarray}

Based on eq. (\ref{f17}) the expression for the tensor $T_{ij}$ has this form:
\begin{eqnarray}
\label{f23}
&&T_{ij}=\frac{T}{9} (g_{ij}+5G_iG_j)
\end{eqnarray}

Combining eq. (\ref{f22}), eq. (\ref{f23}) and constraint on the gravitational vector $G_i$ we will get the final form
for the first set of equations for the problem of pure gravitation:
\begin{eqnarray}
\label{f23a}
&&Q_i:=\lambda_1 ({G_{i;s}}^{;s}-{G_{s;i}}^{;s})+\lambda_3 ({G^s_{;s}}^{;i}-{G_{s;i}}^{;s})+TG_i=0 \\
\label{f23b}
&&T_{ij}=\frac{T}{9} (g_{ij}+5G_iG_j)\nonumber\\
&& G^iG_i=1\nonumber
\end{eqnarray}

We can now address the second set of equations obtained by variation of the Lagrangian $L_G$ by the metric tensor $g^{ij}$,
which we denote as $q_{kl}$.

\begin{eqnarray}
\label{f24}
&& q^{kl}:=\frac{1}{\sqrt{g}}\delta (\sqrt{g}\,L_G)/\delta g_{kl}+T^{kl}=0
\end{eqnarray}

The important element in deriving these equations is first to write the Lagrangian $L_G$ in terms of $P^i_{jk}$
 and $g_{ij}$ and 
then, consider the variation of $g_{ij}$ with variables $P^i_{jk}$ to be fixed. Once the variations are calculated, the
tensor $P^i_{jk}$ can be replaced with its expression through gravitational vector $G_i$ by eq. (\ref{f5a}).

The obtained result will be significantly different than the variation of Lagrangian $L_G$ with $G_i$ being fixed
as it was incorrectly done in the first paper \cite{i20}.

The variation with respect to $g_{kl}$ consists of two terms: 

a) variation with respect to algebraic terms
of $g_{kl}$, which is a straight partial derivative by $g_{kl}$ 
and  

b) variation with respect to partial derivatives $g_{kl,m}$ that are part of Christoffel's symbols
$\Gamma{^i_{jk}}$. The end result here will always have the form of a divergence of a three-index tensor 
(${I^{kls}}_{;s}$) due to the fact 
that Christoffel's symbols are proportional to the first derivatives of metric tensor $g_{ij}$.

The detailed calculations of three different Lagrangians 
\begin{eqnarray}
\label{f24a}
&&L_1=G_{k;l}G^{k;l},\quad L_2=G_{k;l}G^{l;k}\quad and \quad L_3=({G^k}_{;k})^2 \nonumber
\end{eqnarray}
are given in Appendices A, B and C, respectively. 
The final results of these calculations eq.(\ref{fA20}), (\ref{fB20}) and (\ref{fC8a}), are shown below.

\begin{eqnarray}
\label{f25}
&&q^{kl}:=\frac{1}{\sqrt{g}}\delta (\sqrt{g}L_G)/ \delta{g_{kl}}+T^{kl}=0\quad,\,\,where\nonumber\\
&&q_{kl}=\lambda_1[\frac7{18}(G^{m;n}G_{m;n})g_{kl}-{G^s}_{;k}G_{s;l}-\frac{5}{9}{G_k}^{;s}G_{l;s}]+\nonumber\\
&&\quad \,\,+\lambda_2[\frac7{18}(G^{m;n}G_{n;m})g_{kl}-\frac79 G_{k;s}{G^s}_{;l}-\frac79 G_{l;s}{G^s}_{;k}]\nonumber\\
&&\quad \,\, +\lambda_3[\frac7{18}({(G^m{;m})}^2g_{kl}-\frac79 G^s_s G_{k;l}-\frac79 G^s_s G_{l;k}]\nonumber\\
&&\quad \,\,+\lambda_1[\frac12{(G_{s;k}G_l+G_{s;l}G_k)}^{;s}+\frac5{18}{(G_{k;s}G_l+G_{l;s}G_k)}^{;s}\nonumber\\
&&\quad \,\, -\frac12 {(G_{k;l}G_s+G_{l;k}G_s)}^{;s}]\nonumber\\
&&\quad \,\,+\lambda_2[\frac19{(G_{m;n}G^n)}^{;m}g_{kl}+\frac5{18}{(G_{s;k}G_l+G_{s;l}G_k)}^{;s}\nonumber\\
&&\quad \,\,+\frac12{(G_{k;s}G_l+G_{l;s}G_k)}^{;s}-\frac12 {(G_{k;l}G_s+G_{l;k}G_s)}^{;s}]\nonumber\\
&&\quad \,\,+\lambda_3[-\frac89{(G^m_{;m}G^s)}_{;s}g_{kl}+\frac79{(G^m_{;m}G_k)}_{;l}+\frac79{(G^m_{;m}G_l)}_{;k}]\nonumber\\
&&\quad \,\,+\frac19T(g_{kl}+5G_kG_l)]
\end{eqnarray}

The first three brackets of the equation above are the variations by algebraic terms of $g_{kl}$, which are the straight
derivatives of $L_G$ by $g_{kl}$. The trace (contraction with $g^{kl}$) of each of these three brackets is zero,
which is due to the quadratic form of Lagrangian.

The last three brackets are due to the variation with respect to $g_{kl,m}$ (which is a part of Christoffel's symbol 
${\Gamma^i}_{jk}$) and have the form of divergence of a flux.

The system of equations (\ref{f23a}) and (\ref{f25}) is the final set of equations that describes
the gravitational vector field ($G_i$) and the metric  $g_{ij}$ associated with it.

There is one more consideration of a rather general character that has certain mathematical (and perhaps physical)
importance. 

If we write our Lagrangian $L_G\sqrt{g}$ as a function of $P^i_{jk}$ only, it contains only first derivatives terms
 that symbolically can be 
written as $(P')^2$. If we replace $P^i_{jk}$ with $cP^i_{jk}$, where "c" is some constant,
we can observe that due to the definition of $g_{ij}$ through $P^i_{jk}$ (eq. \ref{f1}) we will have
$g_{ij}\,\,->c^2g_{ij}$, $g^{ij}\,\,->c^{-2}g^{ij}$ and $\sqrt{g}\,\,->c^4\sqrt{g}$. Christoffel symbols are unchanged
and for the Lagrangian itself we get
$L_G\,\,->c^2L_G$:
\begin{eqnarray}
\label{f30}
&&S=\int\nolimits L_G(cP^i_{jk;l},g_{ij}(cP^i_{jk}) \sqrt{g(cP^i_{jk})}\, d^4x=\nonumber\\
&&\int\nolimits c^2 L_G(P^i_{jk;l},g_{ij}(P^i_{jk}) \sqrt{g(P^i_{jk})} \,d^4x
\end{eqnarray}
If we consider that the variation of $P^i_{jk}$ is only due to changes of constant "c" 
($\delta P^i_{jk}= P^i_{jk}\,\delta c$), then for requirement $\delta S=0$ we will get:
\begin{eqnarray}
\label{f31}
&&\delta S=\int\nolimits 2c\delta c \,[L_G(P^i_{jk;l},g_{ij}(P^i_{jk})]\,\sqrt{g(P^i_{jk})}\, d^4x=0\nonumber\\
&&or\quad L_G=0.
\end{eqnarray}

In other words, due to the quadratic nature of Lagrangian and due to the definition of $g_{ij}$ through $P^i_{jk}$,
the variational principle yields that the gravitational Lagrangian on the solution is zero: $L_G=0$. 

Using this property we will now show that the equations of motion (\ref{f23a}), (\ref{f25}) 
contain the law of conservation in the form of "the divergence of vector is zero".

First, by contracting $q_{kl}$, eq. (\ref{f25}), we get:
\begin{eqnarray}
\label{f26}
&&q:=q^{kl}g_{kl}=(\lambda_1+\lambda_2)(J-\bar J)-2\lambda_3 \bar J+T=0
\end{eqnarray}
Or taking into account the condition $\lambda_2=-\lambda_1-\lambda_3$ we get
\begin{eqnarray}
\label{f27}
&&T=\lambda_3 (J+\bar J)
\end{eqnarray}

Contracting eq. (\ref{f23a} line 1) with $G^i$ and using the $G^iG_i=1$ we get:
\begin{eqnarray}
\label{f28}
&&G^i[\lambda_1 {G_{i;s}}^{;s}+\lambda_2 {G_{s;i}}^{;s}+\lambda_3 {G^s_{;s}}^{;i}+TG_i]=0\nonumber\\
&&or\quad \lambda_1(-L_1)+\lambda_2(J_2-L_2)+\lambda_3(\bar J-L_3)+T=0\nonumber\\
&&or\quad \lambda_2J+\lambda_3\bar J-L_G+T=0
\end{eqnarray}

And substituting T from eq. (\ref{f27}) and assuming that $L_G=0$ we get:

\begin{eqnarray}
\label{f28a}
&&-\lambda_1 J+2\lambda_3 \bar J=0
\end{eqnarray}
which represents a law of conservation in the form of "the divergence of vector is zero."
\begin{eqnarray}
\label{f32}
&&-\lambda_1J+2\lambda_3\bar J=0\quad or\nonumber\\
&&{(I^k)}_{;k}=0\quad where \quad \nonumber\\
&&I_k=(-\lambda_1G_{k;m}G^m+2\lambda_3G^m_{;m}G_k)
\end{eqnarray}

In the first paper on this subject \cite{i20} we required 
the trace of $T_{ij}$ to be equal zero in order to obtain the law of conservation.
As can be seen from above, the requirement of $L_G=0$ on solution
also leads to the law of conservation.

\vskip 3em
{\underline {Static, spherically symmetric solution.}}
\vskip 1em
Our goal here is to find a static, sphere symmetrical solution for $G_i$ and $g_{ij}$, eq. (\ref{f23a}) and (\ref{f25}).
One can always choose the system coordinate in such a way that $g_{ij}$ is diagonal and $g_{22}=-g_2=-r^2$.
\begin{eqnarray}
\label{f33}
g_{ij}=(g_0,-g_1,-g_2,-g_2sin^2\theta)
\end{eqnarray}
where $g_0$, $g_1$, $g_2$ are all positive functions of the radius.

The only non zero $\Gamma$s are
\begin{eqnarray}
\label{f33a}
&&\Gamma^0_{01}=\frac{g_{0\, ,\,1}}{2g_0} \quad \Gamma^1_{00}=\frac{g_{0\, ,\,1}}{2g_1} \quad 
\Gamma^1_{11}=\frac{g_{1\, ,\,1}}{2g_1}\nonumber\\
&&\Gamma^1_{22}=\Gamma^1_{33}/sin^2 \theta=\frac{g_{2\, ,\,1}}{2g_1} 
\quad \Gamma^2_{21}=\Gamma^3_{31}/sin^2 \theta=\frac{g_{2\, ,\,1}}{2g_2} 
\end{eqnarray}

The vector $G_i$ will have only two components: $G_0$ and $G_1$. It should be stressed that by choosing new time variable
$\bar t=t+\tau(r)$ one can always make $G_1=0$, but then $g_{ij}$ might not be diagonal any more.

The boundary conditions at infinity are such that the gravitational vector $G_i$ is constant (defined only by outside matter)
and the metric is flat. At infinity, the system of coordinates can always be chosen such that $G_0(\infty)=G_B$, 
$G_1(\infty)=0$ and 
$g_{ij}(\infty)= G_B^2 (1, -1, -1, -1)$, so the condition $G^iG_i=1$ holds.
 
In the case of spherical symmetry, we have five unknowns - $g_{00}$, $g_{11}$, $G_0$, $G_1$, T - and a system of 
seven non-zero equations: two for equations $Q_i=0$, eq.(\ref{f23a}) - "0 or t" and "1 or r" components 
and four in $q_{kl}=0$, eq. (\ref{f25}): 
$q_{00}=0$, $q_{22}=q_{33}/sin^2\theta=0$, $q_{11}=0$, $q_{01}=0$ and $G^iG_i=1$.

In our calculations we will be using modified (but equivalent) set of equations.

For the set $Q_i=0$, eq.(\ref{f23a}), instead of $Q_1=0$ we will use the scalar equation obtained by contraction
of $Q_i=0$ with $G^i$ - $G^iQ_i=0$.

For the set $q_{kl}=0$, eq. (\ref{f25}) we will use these two: $q_{00}=0$, $q_{22}=q_{33}/sin^2\theta=0$.
Instead of equation $q_{11}=0$ we will use $q:=q_{ij}g^{ij}=0$. Its equivalence, can be easily seen from
the explicit form of it:
\begin{eqnarray}
\label{f34}
&q=q_{00}g^{00}+q_{11}g^{11}+2q_{22}g^{22}=0
\end{eqnarray}
Since, $q_{00}=0$, $q_{22}=0$ it makes $q=0$ identical to $q_{11}=0$.

And finally instead of equation $q_{01}=0$ we will use $\bar q:=q_{ij}G^iG^j=0$. Its equivalent, 
can be easily seen from its explicit form:
\begin{eqnarray}
\label{f35}
&\bar q=q_{00}(G_0g^{00})^2+q_{11}(G_1g^{11})^2+q_{01}G_0G_1g^{00}g^{11}=0
\end{eqnarray}
Since, $q_{00}=0$, $q_{11}=0$ it makes $\bar q=0$ identical to $q_{01}=0$.

One of the difficulties in presenting the final result here is the algebraic complexity of the formulas involved.
If we simply  write the final expression it will look somewhat ad hoc and it will leave the reader with a choice
to either believe in it or to repeat the calculations on his own. On the other hand, it is equally impossible 
(even in an appendix) to give a full derivation of the results. So the only meaningful alternative here is a 50-50
approach, where the final results are accompanied with some key steps of the calculations, which should guide 
the reader through the derivation process.

If one writes the equations, mentioned few paragraphs above, one finds them to be extremely algebraically lengthy 
and complex, if not unruly. The road to simplification is not obvious and can be quite lengthy, although the final result
is rather simple. Thus in the case of spherical symmetry, the expressions take much simpler forms if expressed
in the new variables defined this way:
\begin{eqnarray}
\label{f36}
&&a)\quad \hat {G_1}=\frac{(G_1)^2 g_0r}{g_1}\nonumber\\
&&b)\quad \hat g=\sqrt{g_0g_1}\nonumber\\
&&c)\quad x=\frac{1}{r}\\
&&d) \,\, the  \,\, derivative \,\,  by \,\, x \,\, is \,\, designated \,\, as \,\, (\,')\nonumber
\end{eqnarray}

Often during calculations, it is convenient as an intermediate step to use normalized $\bar G_0$ and $\bar G_1$ variables 
according to these formulas:
\begin{eqnarray}
\label{f36a}
&&\bar G_0=G_0/\sqrt{g_0} \quad \bar G_1=G_1/\sqrt{g_1}
\end{eqnarray}

The requirement $G^iG_i=1$ will have this form:
\begin{eqnarray}
\label{f36b}
&&(\bar G_0)^2-(\bar G_1)^2=1
\end{eqnarray}

And after differentiating by r we will have:
\begin{eqnarray}
\label{f36c}
&&\bar G_{0,1}\bar G_0=\bar G_{1,1}\bar G_1
\end{eqnarray}

Referring to Appendix D for details of calculations, we now give the expressions for all equations as well as some 
important terms through their covariant components and then through new variables as functions of "x".
\begin{eqnarray}
\label{f37}
&&L_{12}=G_{m;n}G^{m;n}-G_{m;n}G^{n;m}=-x^4(G_0\,')^2\,\, \,eq.\, (\ref{fD5}-\ref{fD6})\\
\label{f37a}
&&L_{32}=G^m_{;m}G^n_{;n}-G_{m;n}G^{n;m}=-\frac{2x^4(\hat G_1)\,'}{\hat g} \quad eq.\, (\ref{fD7}-\ref{fD8})\\
\label{f37b}
&&J=(G^{m;n}G_n)_{;m}=\frac{x^4}{2\sqrt{\hat g}} [({G_0}^2)\,'\frac{1}{\sqrt{\hat g}}] \quad \quad \quad eq.\, (\ref{fD9}-\ref{fD12})\\
\label{f37c}
&&\bar J=(G^m_{;m}G^n)_{;n}=\frac{x^4}{2\sqrt{\hat g}}[({\hat G_1}\,'x-3{\hat G_1})\frac{1}{\sqrt{\hat g}}] ' \quad 
eq.\, (\ref{fD12a}-\ref{fD16})
\end{eqnarray}

We can now write the equations of motion in terms of the new variables.

We begin with the equation  obtained by contracting $q_{kl}$, (\ref{f25}): 
\begin{eqnarray}
\label{f40}
&&q:=q_{kl}g^{kl}=0 \quad or \quad T=\lambda_3(J+\bar J).
\end{eqnarray}
which defines the invariant T. 

The second equation is $q_{22}=0$, (\ref{f25}). While reserving the details of calculations to Appendix E,
 we will write here only the final result. In terms of components, the equation $q_{22}=0$ has this form:
\begin{eqnarray}
\label{f38}
&&q_{22}=0\quad or \quad -\frac{\lambda_1}{2}J+\frac{G_{2;2;s}G^s+G^m_{;m}G_{22}}{g_{22}}=0
\end{eqnarray}

And in terms of new variables:
\begin{eqnarray}
\label{f39}
&&q_{22}=0\quad or \quad -\frac{\lambda_1}{2}J-\frac{\lambda_3x^4}{\sqrt{\hat g}}(\frac{\hat {G_1}}{\sqrt{\hat g}})'=0
\nonumber\\
&&or \quad -\frac{\lambda_1}{2}J-\frac{\lambda_3x^4}{\hat g} ({\hat G_1}'-\frac{G_1{\hat g}'}{2\hat g})=0\nonumber\\
&&or \quad \lambda_1 J=\lambda_3 L_{32}+\frac{\lambda_3G_1{\hat g}'}{\hat g}=0
\end{eqnarray}

As the third equation we will consider a scalar equation $\bar q=0$, obtained by contracting $q_{kl}$ (\ref{f25})
with $G^kG^l$. In terms of components it has this form:
\begin{eqnarray}
\label{f41}
&&\bar q=:q_{kl}G^kG^l=0 \quad or \,\, see \,\, eq.\,\, (\ref{fF13})\nonumber\\
&&\frac{\lambda_1}{2}L_{12}-\frac{\lambda_3}{2}L_{32}+\lambda_1J=0
\end{eqnarray}

Substituting $J$ (\ref{f39}, line 3) in the equation above (\ref{f41}) we will get this expression for the 
third equation:
\begin{eqnarray}
\label{f42}
&&\frac{\lambda_1}{2}L_{12}+\frac{\lambda_3}{2}L_{32}+\frac{\lambda_3G_1{\hat g}'}{\hat g}=0\nonumber\\
&&or \quad \frac{L_G}{2}+\frac{\lambda_3G_1{\hat g}'}{\hat g}=0
\end{eqnarray}

From this expression it follows that if our solution is such that $L_G=0$ (which we require),
then $\hat g=const=G_B^4$ or $g_1=G_B^4/g_0$. If one chooses a system coordinate such that $G_B=1$, we will have 
the same result as in Einstein's GR.

Using the requirement $L_G=0$ (or $\lambda_3L_{32}=-\lambda_1L_{12}$) and the fact that $\hat g =constant$
 we can rewrite the eq. \ref{f41} as:
\begin{eqnarray}
\label{f43}
&&L_{12}+J=0
\end{eqnarray}
Or in variable "x":
\begin{eqnarray}
\label{f44}
&&({G_0})'^2-({G_0}^2)''/2=0\quad or \quad  G_0''=0\nonumber\\
&&with \quad solution \quad  G_0=G_B+K_Tx
\end{eqnarray}

In the expression above, the constant $G_B$ for linear $G_0$ is taken to meet requirements at large distances ($x=0$
or $r=\infty$).
Using expression for $J$ through $G_0$ (eq. \ref{f37b}) we get:
\begin{eqnarray}
\label{f45}
&&J=\frac{K_T^2x^4}{G_B^4}
\end{eqnarray}
From eq. (\ref{f39}) and (\ref{f37a}) we get the expression for $\hat G_1$:
\begin{eqnarray}
\label{f46}
&&\hat {G_1} =G_M-\frac{\lambda_1}{2\lambda_3} {K_T}^2x
\end{eqnarray}
where $G_M$ is  a constant of integration with dimension of 1/length ($cm^{-1}$).
It is important to note that $G_M$ must be positive because at $x=0$ (or $r=\infty$) it represents $( G_1)^2g_0r/g_1$, 
which of course is positive.

And using eq.(\ref{f37c}) we get the expression for $\bar J$;

\begin{eqnarray}
\label{f46a}
&&\bar J=\frac{\lambda_1}{2\lambda_3}\frac{{K_T}^2x^4}{G_B^4}
\end{eqnarray}

Using constraint $G_kG^k=1$ we will get the expression for the $g_0$ component of the metric tensor:
\begin{eqnarray}
\label{f46b}
&&G_kG^k=1\quad or \quad \frac{G_0^2}{g_0}-\frac{G_1^2}{g_1}=1\quad 
or \quad \frac{G_0^2}{g_0}-\frac{\hat G_1 x}{g_0}=1\nonumber\\
&&g_0={G_0}^2-{(\hat G_1)x}\quad\nonumber\\
&&or\quad g_0=G_B^2+(2K_TG_B-G_M)x+(1+\frac{\lambda_1}{2\lambda_3})K_T^2x^2
\end{eqnarray}
And the expression for scalar T defined by eq. (\ref{f40}) has this form:
\begin{eqnarray}
\label{f47}
&&T=\frac{\lambda_3x^4(1+\frac{\lambda_1}{2\lambda_3})K_T^2}{G_B^4}
\end{eqnarray}
The above expressions for $G_0$, $G_1$ (or $\hat G_1$) and $g_{ij}$ as functions of x (or 1/r)
fully define the solution that we were looking for.

We now have to confirm that these solutions satisfy two more equations that we have not yet considered.

The first one is the equation $q_{00}=0$, (\ref{f25}) with k=l=0. 

The second one is  0-component of equation (\ref{f23a}) for vector $G_i$.
\begin{eqnarray}
\label{f47a}
\lambda_1({G_{0;m}}^{;m}-{G_{m;0}}^{;m})+\lambda_3({G^m_{;m}}_{;0}-{G_{m;0}}^{;m})+TG_0=0
\end{eqnarray}

This fact adds no additional information and for the sake of clarity
of this presentation we will do these calculations in Appendix F.

\baselineskip 1pc
\vskip 3em
{\underline {Discussion}}
\vskip 1em
We begin this section by writing the final expressions for $G_i$ and $g_{ij}$ as functions of distance r.

\begin{eqnarray}
\label{f48}
&& G_0=G_B+\frac{K_T}{r}
\end{eqnarray}
\begin{eqnarray}
\label{f49}
&& G_1=\frac{G_B^2}{g_0} \sqrt{\frac{G_M}{r}-\frac{\lambda_1}{2\lambda_3} \frac{K_T^2}{r^2}}
\end{eqnarray}
\begin{eqnarray}
\label{f50}
&&g_0=G_B^2[1-(\frac{G_M}{G_B^2}-\frac{2K_T}{G_B}) \frac{1}{r}+
(1+\frac{\lambda_1}{2\lambda_3})\frac{K_T^2}{G_B^2}\, \frac{1}{r^2}]
\end{eqnarray}
\begin{eqnarray}
\label{f51}
&&g_1=\frac{G_B^4}{g_0}
\end{eqnarray}
\begin{eqnarray}
\label{f52}
&& T_{ij}=\frac{T}{9}(g_{ij}+5G_iG_j)\quad T=\lambda_3(1+\frac{\lambda_1}{2\lambda_3})\frac{K_T^2}{r^4G_B^4}
\end{eqnarray}

The above solutions have four constants: $G_B$, $G_M$, $K_T$ and $\lambda_1/\lambda_3$.
The constant $G_M$ is positive because at large distances it represents a square of $G_1$ (eq. \ref{f49} ).
The constant $K_T$ should be also taken as a positive, because according to eq.(\ref{f48}) it represents the
addition (increase) of the background gravitational field due to the presence of hard matter. 
The value of the constant $\lambda_1/\lambda_3$ is universal - defined by Lagrangian - and,
 as it seems, should be negative so that $G_1$ existed for all r.
It is important to point out that these solutions are exact solutions for a spherically symmetrical gravitational field
and are applicable to any situation on large distances from any concentrated set of masses. 

The value of these constants cannot be defined by a gravitational field alone, but must be derived from a
proper description of a cluster of masses, which we call a point mass. 

In general, the metric $g_{ij}$ differs from the Schwarzschild metric of Einstein's GR by the presence of the quadratic
($1/r^2$) term with proportionality constant $K_T$. This constant is also a proportionality factor 
in the expression for T (or $T_{ij}$ - thus the name $K_T$). If we require that energy-momentum tensor is not zero,
the $K_T$ must be non zero. 

If $K_T$ is small, or at the large distances the quadratic term in $g_0$ 
could be neglected, then the metric becomes a Schwarzschild one with Schwarzschild radius:
\begin{eqnarray}
\label{f59}
&&R_s=\frac{G_M}{G_B^2}
\end{eqnarray}

If $K_T$ is large enough (in terms of $G_M/G_B$), then the metric ($g_0$ component) has no singularity at all. 
It is $G_B^2$ at infinity, then as we move toward the center (r=0) it decreases, reaches its minimal value at some point
and then starts increasing again approaching infinity at r=0. This increase in $G_0$ is equivalent to gravitational repulse.
Of course, it must be kept in mind that this all could be a moot point if the dimensions of point mass 
are bigger than that minimal point.

The conversion of tensor $T^{ij}$ to physical units of energy should be done based on units of dimensions.
In the next formulas we use the sign $\approx$ to indicate dimensions of the physical parameter. For example, 
the vector field $G_i$ has dimensions $cm^{-1}$ and the metric tensor $g_{ij}$ and $T_{ij}$ (or any 2-down-index tensor)
have dimensions $cm^{-2}$.  
\begin{eqnarray}
\label{f59a}
&& T\approx T^i_j \approx 1 \,\,\sqrt{-det(g_{ij})}\approx cm^{-4} \,\, \sqrt{-det(g_{ij})}dx^3
\approx cm^{-1}\nonumber\\
&&\int\nolimits T\,\sqrt{-det(g_{ij})}dx^3\approx \int\nolimits T^0_0\,\sqrt{-det(g_{ij})}dx^3\approx cm^{-1}\nonumber\\
&& \hbar  c \int\nolimits T\,\sqrt{-det(g_{ij})}dx^3\approx mc^2\,\, (or\,\,energy)
\end{eqnarray}
This of course assumes that the Plank constant $\hbar$ is a universal constant.

The constant $K_T$ represents the portion (in terms of $Mc^2$) of energy stored in a gravitational field. 

The physical meaning of constants $G_B$ and $G_M$ becomes very clear if we consider a large (celestial like) body.

The metric tensor $g_{ij}$ (or $g_{00}$) is defined by two constants with dimensions of length: $G_M/G_B^2$ and $K_T/G_B$.
In the point mass approximation, at least from the dimension point of view, both of these constants must be about
the Schwarzschild radius of this point mass ($R_S=MG_n/c^2$), where $G_n$ is the Newton gravitation constant. 
\begin{eqnarray}
\label{f60}
&& \frac{G_M}{G_B^2}\approx \frac{MG_n}{c^2}
\end{eqnarray}

\begin{eqnarray}
\label{f61}
&&\frac{K_T}{G_B}\approx \frac{MG_n}{c^2}
\end{eqnarray}

In addition, the integral of the energy-momentum tensor from the Schwarzschild radius, to infinity
should be equal, with some reasonable factor, to the energy of the point mass - $Mc^2$.

\begin{eqnarray}
\label{f62}
&&\int\nolimits \hbar c T^0_0(4 \pi )\sqrt{-det(g_{ij})}r^2dr \approx K_T^2 \hbar c/R_S\nonumber\\
&&or \quad \approx K_T^2 \hbar c/(MG_n/c^2) \approx Mc^2
\end{eqnarray}
From equation (\ref{f62}) we get:
\begin{eqnarray}
\label{f63}
&&K_T \approx N\sqrt{\frac{G_nM^2}{\hbar c}}=\frac{Nl_p}{l_m}
\end{eqnarray}
where N is the number of particles defined with respect to the proton mass - $M=Nm_p$, $l_p$ is Plank's length and
 $l_m$ is proton's length.
Inserting $K_T$ into equation (\ref{f61}) we will get for $G_B$:
\begin{eqnarray}
\label{f64}
&&G_B \approx \sqrt{\frac{c^3}{G_n \hbar}}=\frac{1}{l_p}
\end{eqnarray}
And using equation (\ref{f60}) we get this expression for $G_M$:
\begin{eqnarray}
\label{f65}
&& G_M \approx R_S*G_B^2 \approx N/(\frac{\hbar c}{mc^2} ) \approx \frac{N}{l_m}
\end{eqnarray}

We now can see that the background concentration ($G_B$) of the gravitational vector field $G_i$ is the 
inverse of Plank's length ($\approx 610^{33} cm^{-1}$). 
The constant $G_M$ is proportional to the number of particles with the dimensional factor
of proton size ($1/lm=m_pc^2/\hbar c$). The constant $K_T$ is also proportional to a number of particles but with  a
factor of $l_p/l_m$.

Another property of this solution is that at large distances it predicts the attraction as the only force
of gravitation. The constant $G_M$ is positive because at large 
distances (r) it corresponds to $G_1^2$. So the law of gravitational attraction is due to the Minkowski
signature of the metric (pseudo-Euclidean space). In contrast, in GR the metric tensor can be chosen to
correspond to either attraction ($g_0=1-r_o/r$) or repulsion ($g_0=1+r_o/r$) and the sign minus is chosen
 to fit Newton's law at large distances.

The energy-momentum tensor of the gravitational field is given by eq. (\ref{f52}). 
It is, like in the case of electromagnetism, inverse to $r^4$. It is worth pointing out that the divergence of $T_{ij}$ 
does not vanish: $T^j_{i;j}\neq 0$.
As we mentioned before, this property ($T^j_{i;j}=0$) does not represent in tensor analysis any law of conservation,
and perhaps should be treated as strictly property of the flat space. 

All solutions ($G_0$, $G_1$ and $T_{ij}$) have singularity at $r=0$. This, as in the flat space field theory, 
is an artifact of our point mass approximation. At small distances r, we cannot neglect the presence of other
parts of tensor potential $P^i_{jk}$ - such as $S_{ijk}$, etc.

There is an interesting question that can be raised: can the constant $K_T$ be negative ($K_T=-\bar K_T,\, \bar K_T>0$)? 
In this case, the value of the time component of the gravitational field ($G_0$) would be decreasing as r approaches 
zero to become zero at the point $r_h=\bar K_T/G_B$. The expectation here is that
both the gravitational field and the metric tensor all turn zero at that point.
This would correspond to a kind of  ether with a "hole" - and $r_h<\bar K_T/G_B$ is meaningless.
The first requirement in this case would be that $\lambda_1/2\lambda_3>0$.
Indeed if we choose $\bar K_T G_B/G_M\lambda_1/2\lambda_3=1$, both $G_0=0$ and $g_0=0$ at $r_h=\bar K_T/G_B$.
However, two problems arise: 1) $G_1=\infty$ at that point and 2) for $g_0$ there is another point - prior to the "hole", 
$r=\bar K_T/G_B (1+\lambda_1/2\lambda_3)>r_h$ - where $g_0$ becomes zero.
Thus we must conclude that $K_T$ is positive.

In the end of this section we would like to point out again that the new equations (\ref{f23a}), (\ref{f25}),
that describe the gravitational field
 and associated with it metric tensor, differ significantly from the Einstein equations of GR ($R_{ij}-\frac{1}{2}
R=T_{ij}$). However, in the case of a static spherical symmetry, they could be made to look close enough to each other.
Indeed, the equation of vector $G_i$ has this form:

\begin{eqnarray}
\label{f67}
&&\lambda_1 ({G_{i;s}}^{;s}-{G_{s;i}}^{;s})+\lambda_3 (G^s_{s;i}-{G_{s;i}}^{;s})+TG_i=0\\
&& or\quad R_{is}G^s=\frac{1}{\lambda_3}(TG_i-\lambda_1 ({G_{i;s}}^{;s}-{G_{s;i}}^{;s})
\end{eqnarray}

In spherical coordinates it can be written as:
\begin{eqnarray}
\label{f68}
&&R^0_0=\frac{1}{\lambda_3}T-\frac{\lambda_1}{\lambda_3} ({G_{0;s}}^{;s}-{G_{s;0}}^{;s})/G_0\\
&&R^1_1=\frac{1}{\lambda_3}T-\frac{\lambda_1}{\lambda_3} ({G_{1;s}}^{;s}-{G_{s;1}}^{;s})/G_1
\end{eqnarray}
The RHS of these equations are proportional to $\frac{1}{r^4}$ - see for example expression for T, eq. (\ref{f52}) -
 and in the first order
of magnitude could be neglected. The remaining equations $R^0_0=0$ and $R^1_1=0$ are equivalent 
for a sphercally symmetric problem to $R_{ij}=0$.

\vskip 2em
\large
{\underline {Gravitational Waves}}
\vskip 1em
The main goal of this section is to demonstrate that
the theory of gravitation proposed in this paper allows existence of gravitational waves. 
To this end, all we need to do is to come up with a wave-like solution. It does not need to be the
complete solution, but just a solution.

It needs to be clarified that
"gravitational waves" does not mean any time dependent behavior of gravitational functions 
($G_i$ and $g_{ij})$. What we mean by a gravitational wave is a solution that exists far away from the
center of mass - so far that the static potential of that mass can be neglected and both the gravitational vector
$G_i$ and the metric tensor $g_{ij}$ could be considered as a small perturbation ($\tilde G_i$, $\tilde g_{ij}$)
over constant gravitational field background and flat metric:
\begin{eqnarray}
\label{GW1}
&&G_i=G^{(o)}_i+\tilde G_i \quad \tilde G_i<<G^{(o)}_i\nonumber\\
&&g_{ij}=g^{(o)}_{ij}+{\tilde g}_{ij}\quad \tilde g_{ij}<<g^{(o)}_{ij}
\end{eqnarray}

where $G^{(o)}_i=(1,0,0,0)$ and $g^{(o)}_{ij}:=\delta_{ij}=Minkowski(1,-1,-1,-1)$.

It is possible to get a covariant linearized expression of equations for $G_i$ (eq. \ref{f25}) and $g_{ij}$ (eq. \ref{f27}),
but in practical terms it does not give much of clarity. The better way is to go directly to time (t),
and Euclidean coordinates ( $x_\alpha$, where $\alpha=1,2,3$ ) representation.
Omitting the lengthy derivations of a desired solution we simply present the solution and demonstrate that
it satisfies all the equations.

\vskip 1em
For a small linear approximation the gravitational field and the metric tensor will satisfy all equations, if
 we have this form:
\begin{eqnarray}
\label{GW2a}
&&\tilde g_{00}=2\tilde G_0\\
\label{GW2b}
&& \tilde g_{0 \alpha}=\tilde G_\alpha\\
\label{GW2c}
&&\tilde g_{\alpha\beta}=0\\
\label{GW2d}
&&\tilde G_i=F_{,i}\\
&&\tilde T=0
\end{eqnarray}
where F(x) is an arbitrary function of coordinates ($t,\,x_\alpha$). 

The first relation, eq.(\ref{GW2a}), comes from the constraint equation for the vector $G_i$ - $G_iG^i=1$:
\begin{eqnarray}
\label{GW3}
&&G_iG^i=1\quad or\quad (G^{(o)}_i+\tilde G_i)(G^{(o)}_j+\tilde G_j)(\delta^{ij}+\tilde g^{ij})=1\nonumber\\
&&(G^{(o)}_iG^{(o)}_j+2G^{(o)}_i\tilde G_j)(\delta^{ij}-\tilde g_{mn}\delta^{im}\delta^{jn})=1\nonumber\\
&&2G^(o)_0\tilde G_0- G^{(o)}_0G^{(o)}_0\tilde g_{00}=0\nonumber\\
&&\quad or\quad \tilde g_{00}=2\tilde G_0
\end{eqnarray}

We now will show that in linearized form the tensor $G_{i;j}$ is zero.
Let us now consider the first derivative of vector $G_i$, $G_{i;j}$
The 00 component of it has this expression:
\begin{eqnarray}
\label{GW4}
&&\tilde {G_{0;0}}=(\tilde {G_{0,0}-\Gamma^k_{00}})=\tilde G_{0,0}-\tilde \Gamma^0_{00}G^{(0)}_0\nonumber\\
&&=\tilde G_{0,0}-\frac{1}{2}(\tilde g_{00,0}+\tilde g_{00,0}-\tilde g_{00,0})\nonumber\\
&&=\tilde G_{0,0}-\frac{1}{2}\tilde g_{00,0}=\quad per\,\, eq.\,\, (\ref{GW2a})\quad =0
\end{eqnarray}

Similarly we can consider the 0$\alpha$ component:
\begin{eqnarray}
\label{GW5}
&&\tilde {G_{0;\alpha}}=(\tilde {G_{0,\alpha}-\Gamma^k_{0\alpha}})=
\tilde G_{0,\alpha}-\tilde \Gamma^0_{0\alpha}G^{(0)}_0\nonumber\\
&&=\tilde G_{0,\alpha}-\frac{1}{2}(\tilde g_{00,\alpha}+\tilde g_{0\alpha,0}-\tilde g_{0\alpha,0})\nonumber\\
&&=\tilde G_{0,\alpha}-\frac{1}{2}\tilde g_{00,\alpha}=\quad per\,\, eq.\,\, (\ref{GW2a})\quad=0
\end{eqnarray}

We now consider the $\alpha$0 component:
\begin{eqnarray}
\label{GW6}
&&\tilde {G_{\alpha;0}}=(\tilde {G_{\alpha,0}-\Gamma^k_{0\alpha}})=
\tilde G_{\alpha,0}-\tilde \Gamma^0_{0\alpha}G^{(0)}_0\nonumber\\
&&=\tilde G_{\alpha,0}-\frac{1}{2}(\tilde g_{00,\alpha}+\tilde g_{0\alpha,0}-\tilde g_{0\alpha,0})=
\tilde G_{\alpha,0}-\frac{1}{2}\tilde g_{00,\alpha}\nonumber\\
&&=per\,eq.\,(\ref{GW2a})=\tilde G_{\alpha,0}-\tilde G_{0,\alpha}\nonumber\\
&&=per\,eq.\,(\ref{GW2d})=F_{,\alpha0}-F_{,0\alpha}=0
\end{eqnarray}

And lastly, we consider the $\alpha \beta$ component:
\begin{eqnarray}
\label{GW7}
&&\tilde {G_{\alpha;\beta}}=(\tilde {G_{\alpha,\beta}-\Gamma^k_{\alpha\beta}})=
\tilde G_{\alpha,\beta}-\tilde \Gamma^0_{\alpha\beta}G^{(0)}_0\nonumber\\
&&=\tilde G_{\alpha,\beta}-\frac{1}{2}(\tilde g_{0\alpha,\beta}+
\tilde g_{0\beta,\alpha}-\tilde g_{\alpha\beta,0})\nonumber\\
&&=per\,eq.\,(\ref{GW2b})=\tilde G_{\alpha,\beta}-\frac{1}{2}\tilde G_{\alpha,\beta}-\frac{1}{2}\tilde G_{\beta,\alpha}
+\frac{1}{2}\tilde g_{\alpha\beta}\nonumber\\
&&=per\,eq.\,(\ref{GW2c})\, and \, eq.\,(\ref{GW2d})=F_{,\alpha\beta}-F_{,\beta\alpha}=0
\end{eqnarray}

Thus we showed that tensor $G_{i;j}$ in linearized terms is identical to zero. The second derivative of gravitational
vector, $G_{i;j;k}$ in first order of approximation is a partial derivative of the tensor $G_{i;j}$
and thus equal to zero as well:
\begin{eqnarray}
\label{GW8}
&& \tilde {G_{i;j;k}}=\tilde {(G_{i;j})}_{,k}=0
\end{eqnarray}

From here it is obvious that every term in eq. (\ref{f23a}, \ref{f25}) is zero and the equations are satisfied.

The solutions (\ref{GW2a} through \ref{GW2d}) show that the metric of the gravitational wave expresses itself
only in time domain, leaving spatial metric flat.

The freedom of function F means that it could be a plane wave moving in the x direction
with a speed V different from the speed of light: F=F(tV-x). 
The linearized equations do not define function F. This, however, does not mean that F can be any function. 
It is possible that the exact solution (or the second order of approximation) will put some restrictions on function F.
For example, it is possible that the exact solution (or the second order of approximation), at least for some constants
$\lambda_1$ and $\lambda_3$, will require that F satisfy an equation similar to the Maxwell equation for vector potential,
which would require the gravitational wave to move with the speed of light.

Also, the trace of the energy-momentum tensor $\tilde T$, and thus the whole tensor $\tilde T_{ij}$, is zero. 
This should not be surprising since the energy-momentum tensor is proportional to the square of
perturbation.

\newpage
\vskip 1em
 {\underline{Cosmology and Ether.}}
\vskip 1em
There are two distinctively different physical scenarios that can be viewed here depending on whether one accepts the idea
of ether.
In the first scenario, based on the existence of ether, the gravitational field exists everywhere with some level of $G_B$
(that is the ether). It is not difficult to show that all equations of the gravitational field (\ref{f23a}) and (\ref{f25})
are satisfied if $g_{ij}$ is a time independent 4-D sphere and $G_0=1$, $G_1=0$.  Indeed, for the metric of the 3-D sphare
of radius R(t) has this form:
\begin{eqnarray}
\label{CM1}
g_{ij}=(dt)^2[1-\dot {R^2}]-R^2[(d\chi)^2+sin^2\chi(d\theta)^2+sin^2\chi sin^2\theta(d\phi)^2]
\end{eqnarray}
For the (00) component of the tensor $G_{k;l}$ ($G_{0;0}$) we have:
\begin{eqnarray}
\label{CM2c}
&&G_{0;0}=G_{0,0}-\Gamma^0_{00}G_0=G_{0,0}-\frac{g_{00,0}}{2g_{00}}=(G_0\sqrt{g_0})_{,0}\sqrt{g_0}=0 \nonumber
\end{eqnarray}
due to the constraint $G_kG^k=1$.
The only components of tensor $G_{kl}$ that are not identically zero are:
\begin{eqnarray}
\label{CM2}
&&G_{1;1}=-\Gamma^0_{11}G_0=\frac{1}{2}\frac{g_{11,0}}{g_0}=\frac{\dot R}{R} \frac{g_{11}}{g_0} \nonumber
\end{eqnarray}
\begin{eqnarray}
\label{CM2a}
&&G_{2;2}=-\Gamma^0_{22}G_0=\frac{1}{2}\frac{g_{22,0}}{g_0}=\frac{\dot R}{R} \frac{g_{22}}{g_0}\nonumber
\end{eqnarray}
\begin{eqnarray}
\label{CM2b}
&&G_{3;3}=-\Gamma^0_{33}G_0=\frac{1}{2}\frac{g_{33,0}}{g_0}=\frac{\dot R}{R} \frac{g_{33}}{g_0}\nonumber
\end{eqnarray}
and they are all proportional to $\dot R=dR/dt$ and thus are zero for $R=const$ in the case of a static universe.

In another words, a static universe with a fixed gravitational field is a solution for our system of equations.
All celestial bodies, including galaxies, should be attracted to each other with the universal gravitational constant
$G_n$. The observation of an expanding and accelerating universe should be explained through other means - for
example, interaction of light with a gravitational background field $G_B$ (ether), as suggested in \cite{i20}.

In the second scenario, we have to start by questioning what would happen if all matter is collected in one point mass?
Or, what is the universe of a single point mass? It is reasonable to expect that such a universe is finite. 
It is also reasonable to assume that $G_B=0$. In this case, the solution for the gravitational field and metric 
has this form: 
\begin{eqnarray}
\label{CM3}
&&G_0=A\frac{1}{r} \quad A>0\nonumber\\
&&\hat G_1=B\frac{1}{r}-A\frac{\lambda_1}{2\lambda_3r^2}\nonumber\\
&&g_{00}=A^2(1+\frac{\lambda_1}{2\lambda_3})\frac{1}{r^2}-B/r=k(R^2_{max}/r^2-R_{max}/r)
\end{eqnarray}

where k and $R_{max}$ (or A, B) are constants defined by mass.

If $-1<(\lambda_1/2\lambda_3)<0$, the gravitational field ($G_0$ and $G_1$) has singularity only at $r=0$, 
which is an artifact of point mass approximation.
The metric has meaning only for $r<R_{max}$. 
The distance from the center of the mass ($r=0$) to the farthest point, where the metric has meaning ($R_{max}$) is finite:
\begin{eqnarray}
\label{CM4}
s=\int \sqrt{g_1}\,dr< \infty.
\end{eqnarray}

The 2D geometrical analogy is a deformed sphere with mass located at the North pole.

The probe mass inside this universe will be repulsed from the star since the $g_{00}$ decreases with distance.
The Newton law (non relativistic limit) or Newton potential ($\phi_n$) will have quite a different form:
\begin{eqnarray}
\label{CM5}
&&\phi_{n}=\Gamma ^r_{tt}=1/4(g^2_{00})=k^2/4(R^2_{max}/r^2-R_{max}/r)^2
\end{eqnarray}

The probe mass starting at some point ($r=r_0$)will be moving away from the center of the mass
 to the farthest point where it stops ($dr/dt=0$). 
In the case of non radial movement, the probe mass probably will be oscillating over the South pole - 
first overshooting and then reversing its movement and going back.

If we try to take into account the gravitational field of the probe mass 
(that is the universe consisting of two masses M and m with $M>>m$) we will find that the gravitational field near 
the small mass is attractive - the mass m has the constant background field of mass M. 
However, as we move away from the small mass, 
the repulsive effect of the big mass M will overcome the decreasing force of attraction to the small mass 
and the gravitational force will change its polarity. 

Is this the situation that governs the movement of galaxies? Could it be that galaxies are so far away from each other
that they pull away from each other? Could there be a situation when a galaxy rips itself away from this universe
and creates its own universe? We are obviously in the beginning of researching this field with many 
more questions than answers.

It is interesting to note that in the framework of General Relativity, one can entertain similar questions of the universe
of a single mass. In other words, what would be the metric if all masses are collected together? 
Though this is not an unreasonable question, for some reasons it has never been addressed in the literature on GR.
The metric solution of Einstein equation $R_{ij}=0$ then must be:
\begin{eqnarray}
\label{CM6}
&&g_{00}=Rs/r. 
\end{eqnarray}

It also has a repulsive force on a probe mass. But, unlike in the theory discussed above,
the Einstein single mass universe is infinite:
\begin{eqnarray}
\label{CM7}
&&s=\int \nolimits \sqrt{g_1} dr \approx r^{\frac{3}{2}}
\end{eqnarray}

In case of radial movement, this would mean that the probe mass will be moving away from the center of the universe
to infinity with its velocity (dr/dt) decreasing as $1/\sqrt(t)$.

\newpage
\baselineskip 1pc
\vskip 3em
{\underline {Conclusion and Comments}}
\vskip 1em

In this section we would like to summarize the results of this paper and to discuss the alternative approaches
for gravitational theory within the framework of "tensor potential description of matter and space".
\vskip 1em

The goal of this paper was to show that there is an alternative to the GR formulation of gravitation - with
 curved space as part of it - that includes the Schwartschild solution for a point mass as an extreme 
case (parameter $K_T=0$).

However, unlike GR, this theory includes gravitational matter represented by the vector $G_i$ with the condition $G^iG_i=1$,
which is strictly derived in this theory. It also contains a clear definition of the gravitational
energy-momentum tensor $T^{ij}$.

The metric of a point mass is a quadratic function of inverse radius (1/r)
with three parameters with clear physical meanings:

a) $G_B$, a background gravitational field defined by outside matter - for example, for our solar system it is the 
gravitational field of our galaxy or global ether.

b) $K_T$, which defines the value (magnitude) of the energy-momentum tensor $T^{ij}$ - eq. (\ref{f52}).

c) $G_M=N/l_m=N\bar h c/{m_p}^2$, the parameter of  inverse length, proportional to the number of particles and 
 reciprocle to the proton (neutron) atomic length.

The Schwarzschild radius, is defined by these three parameters. 

Unlike GR, the presented theory also showed that gravitational attraction is due to the Minkowski signature of space.
 It also gave the indication - due to the quadratic nature of the metric tensor with respect to 1/r - of the possibility
in some cases (small distances) of gravitational repulsion.

The gravitational theory presented here has extra parameters defining the metric tensor and therefore the physics of this
theory is reacher than that presented in GR.
For example, it includes the possibility of black holes (or horizon), 
although due to the presence of matter ($G_i$) the theory might place some restrictions on behavior of the black holes.

Compared to GR, the presented theory has a completely different set of equations describing the motion of the 
metric tensor $g_{ij}$.
These new equations make the description of the gravitational waves and the cosmological problem much more complicated.
We only touch briefly on these subjects, which undoubtedly require much deeper investigations.

\vskip 2em

Building this theory we made several assumptions (postulates):

a) The metric tensor is defined through $P^i_{jk}$ by a quadratic expression.

b) The choice of $\gamma$s is taken to be $\gamma_1=-18$, $\gamma_2=12$, $\gamma_3=1$.

c) The separation of matter into seven sub-fields is based on the symmetry of the tensor $P_{ijk}$.

d) The choice of vector $G_i$ as a gravitational field.

e) The "independece of sub-fields" postulate.

\vskip 1em
In choosing these postulates we tried to use a principle of "uniqueness of choice", requiring an existence of 
only one possible result stemming from the postulate.

There is only one way (quadratic form) that the metric tensor can be defined through $P^i_{jk}$.
There is only one vector field (for given $\gamma$s) 
that can by itself form space. 
There is only one form of interaction (through the metric tensor) between sub-fields of matter.
There is one simple way for the energy-momentum tensor to be incorporated in this theory.

\vskip 1em
The gravitational theory presented here is derived from a general theory for the description of matter and
space using the 3-index tensor potential proposed in \cite{i20}. 
This description is a natural extension into curved space of the flat space approach
 that has been (and still is) the work horse of theoretical physics for many years.

From another point of view, this formalism is an old Eddington's idea taken in reverse. 
Instead of starting with affine connections and then separating them into space ($g_{ij}$) and matter ($P^i_{jk}$),
we start with matter ($P^i_{jk}$) and then use it to define space through its metric tensor ($g_{ij}$) by eq (\ref{f1}).
Philosophically, instead of geometrization of physics, the new theory advances an idea of materialization of space. 
In fact, one of the main reasons for constructing the theory in a this way was a desire to avoid empty space, be it flat or 
curved.

The approach presented here is not unique. There are other possibilities in constructing the theory
 of "tensor potential description of matter and space".

If we drop the "independence of sub-fields" postulate, we can find a Lagrangian (in fact probably several of them)
in such a way that $T^{ij}=\hat T(g^{ij}-4G^iG^j)$. This energy-momentum tensor, just like the one for 
electro-magnetic field,
has zero trace ($T=T^i_i=0$) and $\hat T$ is defined by the second order invariant ($\hat T=T^{ij}T_{ij}/12$).
In this case, the equations of motion $Q_{ijk}=0$ for $P^i_{jk}$,
 ($Q^{jk}_i:=\delta L_M/\delta P^i_{jk}=0$) are not fully symmetrical, but 
still lead to a vector equation similar to eq. (\ref{f23a}). 
Of course, the equation for $g_{ij}$ ($q^{ij}:=\delta L_M/\delta g^{jk}=0$) will have a completely different form than 
eq.(\ref{f25}).  The work of this paper would have to be repeated for those Lagrangians to see if they yield a
Schwartschild type solution. The difficulty will then multiply when we address the other fields 
(such as the electromagnetic field). It would require that the equations of motion for $P^i_{jk}$ contain
an acceptable solution when the matter $P^i_{jk}$ consists of only
gravitation and electromagnetic sub-fields. 
It is not difficult to show that a general form  of the Lagrangian containing only squares of $P^i_{jk;l}$
has 40 terms (40 $\lambda$s constants) instead of three as in eq. (\ref{f11}).
This considerably increases the difficulty of finding the right set of $\lambda$s. 
Perhaps this work must be done before we can declare that the "independece of sub-fields" postulate
is necessary. Redardless, it would make much more practical sense to first investigate the case with
the "independece of sub-fields" postulate.

There was another very atractive possibility. Having $P^i_{jk}$ and metric $g_{ij}$ we can go back to Eddington's 
affine connections $\Gamma^i_{jk}$ and from them to Riemann tensor and then to a Lagrangian proportional to square 
of Riemann tensor.
This would be identical to Hehl's metric-affine (MAG) theory with the following definition of the metric tensor:
\begin{eqnarray}
&&g_{kl}=K_{kl}[\Gamma^i_{jk}-\frac{1}{2}g^{is}(g_{is,j}+g_{js,i}-g_{ij,s})]
\end{eqnarray}
with an additional term for a Lagrangian 
($L_T$) corresponding to a constraint defining metric tensor $g_{ij}$:

\begin{eqnarray}
&&L_T=Z^{kl} \{ g_{kl}-K_{kl}[ (\Gamma^i_{jk}-\frac{1}{2}g^{is}(g_{is,j}+g_{js,i}-g_{ij,s})] \}
\end{eqnarray}
where $K_{kl}$ is a quadratic function of eq. (\ref{f1}) and $Z^{kl}$ are the Lagrange coefficients.

This approach might be worthy of consideration, but it was not our first choice for the following reasons:

a) It has an implicit definition for the metric tensor - both RHS and LHS include the metric tensor.

b) The energy-momentum tensor, defined as $\delta (\sqrt{g} L)/ \delta g_{kl}\,\sqrt{g}$, 
and the Lagrangian coefficients $Z^{kl}$ do not 
have a simple relationship between themselves. In our approach those parameters are the same.

c) The separation of matter ($P^i_{jk}$) into sub-fields (gravitational, electromagnetic, etc.) is not obvious.
Should it be done on the basis of the tensor potential $P^i_{jk}$ or on the basis of the Riemann tensor? 
For example, should the gravitational field be defined through vector potential derived from $P^i_{jk}$ as in (\ref{f4}),
or should it be proportional to the Ricci tensor constructed from $R^i_{jkl}$?

\vskip 1em
In conclusion, we would like to mention that the result of this paper is important because it
demonstrates that in the framework of "tensor potential description of matter and space"
there is a derivation of the theory of gravitation. In \cite{i20} we outlined the general idea, 
but in this paper we did the first practicle application of this theory toward one physical matter - gravitation.

\newpage
{\it \underline{Appendix A:}}
\vskip 0.5em
\vskip 1em

The goal of this appendix is to derive the equation $q_{kl}=0$ , where $q_{kl}$
 is a variation of the Lagrangian $L_{(G)}$ with respect to $g_{kl}$. In this appendix we will do the calculations for 
the Lagrangian $L_1=G_{k;l}G^{k;l}$.

We begin by writing the Lagrangian as a function of $P^i_{jk}$:
\begin{eqnarray}
\label{fA1a}
&&G_k=\frac{1}{3}(2{\bar P}^i_{jk}g^j_i+{\bar P}^m_{st}g_{mk}g^{st})\\
\label{fA1b}
&&G_{k;l}=\frac{1}{3}(2{\bar P}^i_{jk;l}g^j_i+{\bar P}^m_{st;l}g_{mk}g^{st})\\
\label{fA1c}
&&{\bar P}_{ijk}=\frac{1}{6}(G_ig_{jk}+G_jg_{ik}+G_kg_{ij})\\
\label{fA1d}
&&L_1=G_{k;l}(\bar P^i_{jk})G^{k;l}(P^i_{jk})
\end{eqnarray}
Here $\bar P^i_{jk}$ is the symmetric part in low indices of the $P^i_{jk}$ tensor and
 $g^i_j$ are Kronneker symbols ($g^i_j=1$, if $i=j$ and $g^i_j=0$ otherwise).

The variation of $L_1$ with respect to $g_{kl}$ consists of two parts. The first part is with respect to algebraic terms
 (including $\sqrt{g}$), which we denote as $q_{(g)}^{kl}$. The second part is with respect to the partial derivative
 of $g_{ij}$ ($g_{ij,k}$)
 that are part of Christoffel symbols  ${\Gamma}^i_{jk}$, which we denote as $q_{(\Gamma)}^{kl}$.

The $q_{(g)}^{kl}$ is a straight partial derivative of $L_1$ by $g_{kl}$:
\begin{eqnarray}
\label{fA2}
&&q_{(g)}^{kl}=\frac{\partial (L_1\sqrt{g})}{\sqrt{g}\partial g_{kl}}\nonumber\\
&&=\frac{\partial L_1(P^i_{jk})}{\partial g_{kl}}+\frac{1}{2}L_1g^{kl}
\end{eqnarray}
where
\begin{eqnarray}
\label{fA2a}
&&\frac{\partial L_1}{\partial g_{kl}}=
\partial[ (\frac{2}{3}{\bar P}^i_{jm;n}g^j_i+\frac{1}{3}{\bar P}^v_{st;n}g_{vm}g^{st})\nonumber\\
&&\quad \quad \quad (\frac{2}{3}{\bar P}^i_{ja;b}g^j_i+\frac{1}{3}
{\bar P}^v_{st;b}g_{va}g^{st})g^{ma}g^{nb}]/\partial g_{kl}
\end{eqnarray}
After opening the parentheses we get:

\begin{eqnarray}
\label{fA3}
&&=\partial (\frac{4}{9} {\bar P}^i_{jm;n}{\bar P}^u_{va;b}g^j_ig^v_ug^{ma}g^{nb})/\partial g_{kl}\nonumber\\
&&+\partial (\frac{4}{9} {\bar P}^i_{jm;n}g^j_i{\bar P}^m_{st;b}g^{st}g^{nb})/\partial g_{kl}\nonumber\\
&&+\partial (\frac{1}{9}{\bar P}^v_{st;m}{\bar P}^u_{xy;n}g_{uv}g^{st}g^{xy}g^{mn})/\partial g_{kl}
\end{eqnarray}
Using the expression $\partial A_{..x..y..}g^{xy}/\partial g_{kl}=-A_{..x..y..}g^{xk}g^{yl}$ we get:
\begin{eqnarray}
\label{fA4}
&&=(-\frac{4}{9} {\bar P}^i_{js;n}{\bar P}^u_{vt;b}g^j_ig^v_ug^{nb}g^{sk}g^{tl}-
\frac{4}{9} {\bar P}^i_{jm;s}{\bar P}^u_{va;t}g^j_ig^v_ug^{ma}g^{sk}g^{tl})\nonumber\\
&&+(-\frac{4}{9} {\bar P}^i_{jm;n}g^j_i{\bar P}^m_{st;a}g^{sk}g^{tl}g^{na}
-\frac{4}{9} {\bar P}^i_{jm;n}g^j_i{\bar P}^m_{st;a}g^{st}g^{nk}g^{al})\\
&&+(\frac{1}{9}{\bar P}^k_{st;m}{\bar P}^l_{xy;n}g^{st}g^{xy}g^{mn}
-\frac{1}{9}{\bar P}^v_{st;m}{\bar P}^u_{xy;n}g_{uv}g^{sk}g^{tl}g^{xy}g^{mn}\nonumber\\
&&-\frac{1}{9}{\bar P}^v_{st;m}{\bar P}^u_{xy;n}g_{uv}g^{st}g^{xk}g^{yl}g^{mn}-
\frac{1}{9}{\bar P}^v_{st;m}{\bar P}^u_{xy;n}g_{uv}g^{st}g^{xy}g^{mk}g^{nl})\nonumber
\end{eqnarray}
Replacing the tensor $bar P^i_{jk}$ with its expression through $G_i$ and taking into consideration that
 any contraction of $P^i_{jk}$ is the vector $G_i$, we get:
\begin{eqnarray}
\label{fA5}
&&=(-\frac{4}{9}G^{k;n}G^{l;b}g_{nb}-\frac{4}{9}G^{m;k}G^{a;l}g_{ma})+\nonumber\\
&&(-\frac{4}{9}G_{m;n}{\bar P}^{mkl;n}-\frac{4}{9}G^{m;k}G^{n;l}g_{mn})+\nonumber\\
&&+(\frac{1}{9}G^{k;m}G^{l;n}g_{mn}-\frac{2}{9}G_{m;n}\bar P^{mkl;n}-\frac{1}{9}G^{m;k}G^{n;l}g_{mn})
\end{eqnarray}
Combining similar terms and symmetrisizing by indices k and l:
\begin{eqnarray}
\label{fA6}
&&=-\frac{3}{9}G^{k;m}G^{l;n}g_{mn}-G^{m;k}G^{n;l}g_{mn}-\frac{6}{9}G_{m;n} \bar P^{mkl;n}
\end{eqnarray}
Replacing $\bar P^{mkl;n}$ of the expression above with its expression through $G_i$
 using eq. (\ref{fA1c}) we get:
\begin{eqnarray}
\label{fA7}
&&=-\frac{3}{9}G^{k;m}G^{l;n}g_{mn}-G^{m;k}G^{n;l}g_{mn}\nonumber\\
&&-\frac{1}{9}G_{m;n}(G^{m;n}g^{kl}+G^{k;n}g^{lm}+G^{l;n}g^{mk})
\end{eqnarray}
and after opening the parentheses:
\begin{eqnarray}
\label{fA8}
&&-\frac{5}{9}G^{k;m}G^{l;n}g_{mn}-G^{m;k}G^{n;l}g_{mn}-\frac{1}{9}G_{m;n}G^{m;n}g^{kl}
\end{eqnarray}
Since $g_{kl}$ is symmetric in low indices, so should be $q_{kl}$ - which it is.

Taking into account eq. (\ref{fA2}) and eq. (\ref{fA8}) we get the final result for $q_{(g)}^{kl}$:
\begin{eqnarray}
\label{fA9}
&&q_{(g)}^{kl}=\frac{\partial (L_1\sqrt{g})}{\sqrt{g}\partial g_{kl}}\nonumber\\
&&=-\frac{5}{9}G^{k;m}G^{l;n}g_{mn}-G^{m;k}G^{n;l}g_{mn}+\frac{7}{18}G_{m;n}G^{m;n}g^{kl}
\end{eqnarray}
As a check point, the contraction with $g_{kl}$ should be zero - as it is. 
This is due to the quadratic form of the Lagrangian with respect to the metric tensor $g^{ij}$.

\vskip 2em
We now derive the expression for the variation of the Lagrangian $L_1$ ($L_1=G^{k;l}G_{k;l}$) with respect to partial 
derivatives of $g_{ij}$ ($g_{ij,k}$), which are part of the Christoffel symbols  ${\Gamma}^i_{jk}$ - denoted
 as $q_{(\Gamma)}^{kl}$. 
The simplest way of doing this is to calculate the deviation $\delta_{(\Gamma)} L_1$. 
Again, as above, we start with writing $L_1$ as a function of $P^i_{jk}$.

\begin{eqnarray}
\label{fA10}
&&\delta_{(\Gamma)} L_1=2G^{k;l}\delta_{(\Gamma)} G_{k;l}=2G^{k;l}\frac{1}{3}\delta_{(\Gamma)}
 (2{\bar P}^i_{jk;l}g^j_i+{\bar P}^m_{ij;l}g^{ij}g_{mk})\nonumber\\
&&=G^{k;l}\frac{2}{3}(2g^j_i\delta_{(\Gamma)}{\bar P}^i_{jk;l}+g^{ij}g_{mk}\delta_{(\Gamma)} {\bar P}^m_{ij;l})
\end{eqnarray}
We now insert the explicit expressions for $\bar P^i_{jk;l}$ through the Christoffel symbols:
\begin{eqnarray}
\label{fA11}
&&=G^{k;l}\frac{2}{3}[2g^j_i\delta_{(\Gamma)}({\bar P}^i_{jk\,,\,l}+\Gamma^i_{ls}{\bar P}^s_{jk}
-\Gamma^s_{lj}{\bar P}^i_{sk}-\Gamma^s_{lk}{\bar P}^i_{js})\nonumber\\
&&+g^{ij}g_{mk}\delta_{(\Gamma)}({\bar P}^m_{ij\,,\,l}+\Gamma^m_{ls}{\bar P}^s_{ij}-\Gamma^s_{li}
{\bar P}^m_{sj}-\Gamma^s_{lj}{\bar P}^m_{is})]\nonumber\\
&&=G^{k;l}\frac{2}{3}[2g^j_i(\delta\Gamma^i_{ls}\,{\bar P}^s_{jk}-\delta\Gamma^s_{lj}\,
{\bar P}^i_{sk}-\delta\Gamma^s_{lk}{\bar P}^i_{js})+\nonumber\\
&&\quad g^{ij}g_{mk}(\delta\Gamma^m_{ls}\,{\bar P}^s_{ij}-\delta\Gamma^s_{li}\,
{\bar P}^m_{sj}-\delta\Gamma^s_{lj}\,{\bar P}^m_{is})]
\end{eqnarray}
In the last expression we used the fact that $\delta_{\Gamma}{\bar P}^i_{jk\,,\,l}=0$
 since the term $\partial P^i_{jk}/\partial x^l$ does not include $g_{ij}$.
Combining similar terms and contracting $\bar P^i_{jk}$ where it is possible back to $G_i$ we get:
\begin{eqnarray}
\label{fA12}
&&=G^{k;l}\frac{2}{3}\,(-2G_s\,\delta\Gamma^s_{lk}+g_{mk}G^s\,\delta\Gamma^m_{ls}-
2g_{mk}g^{ij}{\bar P}^m_{sj}\,\delta\Gamma^s_{li})
\end{eqnarray}

We now transition the Christoffel symbols of the second type ($\Gamma^i_{jk}$) to the Christoffel symbols 
of the first type
$\Gamma^i_{jk}=g^{im}\Gamma_{mjk}$. The variation of the term $g^{im}$ in front of $\Gamma$ could be neglected
 since the end result is proportional to $\Gamma$s and will vanish when transfered to the covariant form.

\begin{eqnarray}
\label{fA13}
&&=-\frac{4}{3}G^{k;l}G^s\delta\Gamma_{slk}+\frac{2}{3}G^{k;l}G^s\delta\Gamma_{kls}-
\frac{4}{3}G^{k;l}\bar P^{ksi}\delta\Gamma_{sli}
\end{eqnarray}
Substituting the explicit expression for $\Gamma_{ijk}$ through $g_{ij\,,\,m}$ we get:
\begin{eqnarray}
\label{fA14}
&&=(-\frac{4}{3}G^{k;l}G^s)\frac{1}{2}(\delta g_{sk\,,\,l}+\delta g_{sl\,,\,k}-\delta g_{kl\,,\,s})\nonumber\\
&&+(\frac{2}{3}G^{k;l}G^s)\frac{1}{2}(\delta g_{kl\,,\,s}+\delta g_{ks\,,\,l}-\delta g_{ls\,,\,k})\nonumber\\
&&-(\frac{4}{3}G^{k;l}g_{mk}{\bar P}^{msi})\frac{1}{2}(\delta g_{sl\,,\,i}+\delta g_{si\,,\,l}-\delta g_{li\,,\,s})
\end{eqnarray}

The three terms of $g_{ij\,,\,l}$ in each parentheses could be written as one after renaming indices and combining similar
terms. Also note that $\bar P^{msi}$ is fully symmetrical and thus the first and the third terms in the last parentheses
cancel each other.
\begin{eqnarray}
\label{fA15}
&&=-\frac{2}{3}(G^{k;s}G^l+G^{s;l}G^k-G^{k;l}G^s)\delta g_{kl,s}\nonumber\\
&&+\frac{1}{3}(G^{k;l}G^s+G^{k;s}G^l-G^{s;l}G^k)\delta g_{kl,s}\nonumber\\
&&-(\frac{2}{3}G^{n;s}g_{mn}{\bar P}^{mkl})\frac{1}{2}\delta g_{kl\,,\,s}
\end{eqnarray}

Substituting $\bar P^{mkl}$ through vector $G_i$ we get:
\begin{eqnarray}
\label{fA16}
&&=-\frac{2}{3}(G^{k;s}G^l+G^{s;l}G^k-G^{k;l}G^s)\delta g_{kl,s}\nonumber\\
&&+\frac{1}{3}(G^{k;l}G^s+G^{k;s}G^l-G^{s;l}G^k)\delta g_{kl,s}\nonumber\\
&&-(\frac{2}{3}G^{n;s}g_{mn})\frac{1}{6}(\underline{G^mg^{kl}}+G^kg^{ml}+G^lg^{km})\delta g_{kl,s}
\end{eqnarray}

The underlined term vanishes due to the constraint $G^kG_k=1$ - $G^{k;l}G_k=1/2(G^kG_k)^{;l}=0$.
We now apply the "partial integration rule" of the variational method and transfer $\partial_s$ from the metric
tensor to the term in front of it with the opposite sign:

\begin{eqnarray}
\label{fA17}
&&=-[-\frac{2}{3}(G^{k;s}G^l+G^{s;l}G^k-G^{k;l}G^s)]_{\,,\,s}\,\delta g_{kl}\nonumber\\
&&-[+\frac{1}{3}(G^{k;l}G^s+G^{k;s}G^l-G^{s;l}G^k)]_{\,,\,s}\,\delta g_{kl}\nonumber\\
&&-[-(\frac{2}{3}G^{n;s}g_{mn})\frac{1}{6}(G^kg^{ml}+G^lg^{km})_{\,,\,s}]\,\delta g_{kl}
\end{eqnarray}

We can now replace the partial derivatives with the covariant ones. Tensor $\delta g_{kl}$ is symmetrical and so should be
the expression in the front of it. After symmetrization we get:

\begin{eqnarray}
\label{fA18}
&&=\{\frac{1}{3}[(G^{s;k}G^l+G^{s;l}G^k)+(G^{k;s}G^l+G^{l;s}G^k)-(G^{k;l}G^s+G^{l;k}G^s)]\nonumber\\
&&+\frac{1}{6}[(G^{s;k}G^l-G^{s;l}G^k)+(G^{k;s}G^l+G^{l;s}G^k)-(G^{k;l}G^s+G^{l;k}G^s)]\nonumber\\
&&+\frac{1}{9}[(G^{k;s}G^l+G^{l;s}G^k)]\}_{;s}\delta g_{kl}
\end{eqnarray}
And collecting similar terms we get:

\begin{eqnarray}
\label{fA18a}
&&=[\frac{1}{2}(G^{s;k}G^l+G^{s;l}G^k)+\frac{5}{18}(G^{k;s}G^l+G^{l;s}G^k)\nonumber\\
&&-\frac{1}{2}(G^{k;l}G^s+G^{l;k}G^s)]_{;s}\delta g_{kl}
\end{eqnarray}
From which it follows that the final expression for $q^{kl}_{(\Gamma)}$ is:

\begin{eqnarray}
\label{fA19}
&&q_{\Gamma}^{kl}:=\delta_{(\Gamma)} L_1/\delta g_{kl}=\nonumber\\
&&=[\frac{1}{2}(G^{s;k}G^l+G^{s;l}G^k)+\frac{5}{18}(G^{k;s}G^l+G^{l;s}G^k)\nonumber\\
&&\quad \quad -\frac{1}{2}(G^{k;l}G^s+G^{l;k}G^s)]_{;s}
\end{eqnarray}

Combining the expression for $q_{g}^{kl}$, eq. (\ref{fA9}), and the term for $q_{\Gamma}^{kl}$, eq. (\ref{fA19}),
 we get the final result for the
variation of the Lagrangian $L_1=G^{k;l}G_{k;l}$ with respect to $g_{kl}$:
\begin{eqnarray}
\label{fA20}
&&q^{kl}=\frac{\partial (L_1\sqrt{g})}{\sqrt{g}\partial g_{kl}}\nonumber\\
&&=-\frac{5}{9}G^{k;m}G^{l;n}g_{mn}-G^{m;k}G^{n;l}g_{mn}+\frac{7}{18}G_{m;n}G^{m;n}g^{kl}+\nonumber\\
&&=[\frac{1}{2}(G^{s;k}G^l+G^{s;l}G^k)+\frac{5}{18}(G^{k;s}G^l+G^{l;s}G^k)\nonumber\\
&&-\frac{1}{2}(G^{k;l}G^s+G^{l;k}G^s)]_{;s}
\end{eqnarray}

\newpage
{\it \underline{Appendix B:}}
\vskip 0.5em
The goal of this appendix is to derive the equation $q_{kl}=0$ , where $q_{kl}$
 is a variation of the Lagrangian $L_{(G)}$ with respect to $g_{kl}$. 
In this appendix we will do the calculations for 
the Lagrangian $L_2=G_{l;k}G^{k;l}$

We begin by writing the Lagrangian as a function of $P^i_{jk}$:
\begin{eqnarray}
\label{fB1a}
&&G_k=\frac{1}{3}(2{\bar P}^i_{jk}g^j_i+{\bar P}^m_{st}g_{mk}g^{st})\\
\label{fB1b}
&&and\quad G_{k;l}=\frac{1}{3}(2{\bar P}^i_{jk;l}g^j_i+{\bar P}^m_{st;l}g_{mk}g^{st})\\
\label{fB1c}
&&{\bar P}_{ijk}=\frac{1}{6}(G_ig_{jk}+G_jg_{ik}+G_kg_{ij})\\
\label{fB1d}
&&L_2=G_{l;k}(P^i_{jk})G^{k;l}(P^i_{jk})
\end{eqnarray}
Here $\bar P^i_{jk}$ is the symmetric part in low indices of the $P^i_{jk}$ tensor and
 $g^i_j$ are Kronneker symbols ($g^i_j=1$, if $i=j$ and $g^i_j=0$ otherwise).

The variation of $L_2$ with respect to $g_{kl}$ consists of two parts. The first part is with respect to algebraic terms
 (including $\sqrt{g}$), which we denote as $q_{(g)}^{kl}$. The second part is with respect to the partial derivative
 of $g_{ij}$ ($g_{ij,k}$)
 that are part of Christoffel symbols  ${\Gamma}^i_{jk}$, which we denote as $q_{(\Gamma)}^{kl}$.
The $q_{(g)}^{kl}$ is a straight partial derivative of $L_1$ by $g_{kl}$:

\begin{eqnarray}
\label{fB2}
&&q_{(g)}^{kl}=\frac{\partial (L_2\sqrt{g})}{\sqrt{g}\partial g_{kl}}\nonumber\\
&&=\frac{\partial L_2(P^i_{jk})}{\partial g_{kl}}+\frac{1}{2}L_2g^{kl}
\end{eqnarray}
where
\begin{eqnarray}
\label{fB2a}
&&\frac{\partial L_2}{\partial g_{kl}}=
\partial[ (\frac{2}{3}{\bar P}^i_{jn;m}g^j_i+\frac{1}{3}{\bar P}^v_{st;m}g_{vn}g^{st})\nonumber\\
&&\quad \quad \quad (\frac{2}{3}{\bar P}^i_{ja;b}g^j_i+\frac{1}{3}
{\bar P}^v_{st;b}g_{va}g^{st})g^{ma}g^{nb}]/\partial g_{kl}
\end{eqnarray}
and after opening parentheses:
\begin{eqnarray}
\label{fB3}
&&=\partial (\frac{4}{9} {\bar P}^i_{jn;m}{\bar P}^u_{va;b}g^j_ig^v_ug^{ma}g^{nb})/\partial g_{kl}\nonumber\\
&&+\partial (\frac{4}{9} {\bar P}^i_{jn;m}g^j_i{\bar P}^m_{st;b}g^{st}g^{nb})/\partial g_{kl}\nonumber\\
&&+\partial (\frac{1}{9}{\bar P}^n_{st;m}{\bar P}^m_{xy;n}g^{st}g^{xy})/\partial g_{kl}
\end{eqnarray}
Using the expression $\partial A_{..x..y..}g^{xy}/\partial g_{kl}=-A_{..x..y..}g^{xk}g^{yl}$ we get:
\begin{eqnarray}
\label{fB4}
&&=(-\frac{4}{9} {\bar P}^i_{jn;m}{\bar P}^u_{va;b}g^j_ig^v_ug^{nb}g^{mk}g^{al}-
\frac{4}{9} {\bar P}^i_{jn;m}{\bar P}^u_{va;b}g^j_ig^v_ug^{ma}g^{nk}g^{bl})\nonumber\\
&&+(-\frac{4}{9} {\bar P}^i_{jn;m}g^j_i{\bar P}^m_{st;b}g^{sk}g^{tl}g^{nb}
-\frac{4}{9} {\bar P}^i_{jn;m}g^j_i{\bar P}^m_{st;b}g^{st}g^{nk}g^{bl})\nonumber\\
&&+(-\frac{1}{9}{\bar P}^n_{st;m}{\bar P}^m_{xy;n}g^{sk}g^{tl}g^{xy}
-\frac{1}{9}{\bar P}^n_{st;m}{\bar P}^m_{xy;n}g^{st}g^{xk}g^{yl})
\end{eqnarray}
Replacing the tensor $\bar P^i_{jk}$ with its expression through $G_i$, eq. (\ref{fB1c}) and 
taking into consideration that
 any contraction of $\bar P^i_{jk}$ is the vector $G_i$, we get:
\begin{eqnarray}
\label{fB5}
&&=(-\frac{4}{9}G^{k;n}G^{b;l}g_{nb}-\frac{4}{9}G^{l;n}G^{b;k}g_{nb})+\nonumber\\
&&(-\frac{4}{9}G_{n;m}{\bar P}^{mkl;n}-\frac{4}{9}G^{k;n}G^{b;l}g_{nb})+\nonumber\\
&&+(-\frac{2}{9}G_{m;n}\bar P^{nkl;m})
\end{eqnarray}

Combining similar terms and symmetrisizing by indices k and l:
\begin{eqnarray}
\label{fB6}
&&=-\frac{6}{9}(G^{k;m}G^{n;l}g_{mn}+G^{l;m}G^{n;k}g_{mn})-\frac{6}{9}G_{n;m} \bar P^{mkl;n}
\end{eqnarray}
Replacing $\bar P^{mkl;n}$ of the expression above with its expression through $G_i$ using eq. (\ref{fB1c}):
\begin{eqnarray}
\label{fB7}
&&=-\frac{6}{9}(G^{k;m}G^{n;l}g_{mn}+G^{l;m}G^{n;k}g_{mn})\nonumber\\
&&-\frac{1}{9}G_{n;m}G^{m;n}g^{kl}-\frac{1}{9}(G^{k;m}G^{n;l}g_{mn}+G^{l;m}G^{n;k}g_{mn})
\end{eqnarray}
or
\begin{eqnarray}
\label{fB8}
&&=-\frac{7}{9}(G^{k;m}G^{n;l}g_{mn}+G^{l;m}G^{n;k}g_{mn})-\frac{1}{9}G_{n;m}G^{m;n}g^{kl}
\end{eqnarray}
Since $g_{kl}$ is symmetric in low indices, so should be $q_{kl}$ - which it is.

Taking into account (eq. \ref{fB2}) and (eq. \ref{fB8}) we get the final result for $q_{(g)}^{kl}$:
\begin{eqnarray}
\label{fB9}
&&q_{(g)}^{kl}=-\frac{7}{9}(G^{k;m}G^{n;l}g_{mn}+G^{l;m}G^{n;k}g_{mn})\nonumber\\
&&+\frac{7}{18}G_{n;m}G^{m;n}g^{kl}
\end{eqnarray}

As a check point, the contraction with $g_{kl}$ should be zero - as it is. 
this is due to the quadratic form of the Lagrangian with respect to the metric tensor $g^{ij}$.

\vskip 3em
We now derive the expression for the variation of the Lagrangian $L_2$ ($L_2=G^{k;l}G_{l;k}$) with respect to partial 
derivatives of $g_{ij}$ ($g_{ij,k}$), which are part of the Christoffel symbols  ${\Gamma}^i_{jk}$ - denoted
 as $q_{(\Gamma)}^{kl}$. 
The simplest way of doing this is to calculate the deviation $\delta_{(\Gamma)} L_2$. 
Again, as above, we start with writing $L_2$ as a function of $P^i_{jk}$.

\begin{eqnarray}
\label{fB10}
&&\delta_{(\Gamma)} L_2=2G^{l;k}\delta_{(\Gamma)} G_{k;l}=2G^{l;k}\frac{1}{3}\delta_{(\Gamma)}
 (2{\bar P}^i_{jk;l}g^j_i+{\bar P}^m_{ij;l}g^{ij}g_{mk})\nonumber\\
&&=G^{l;k}\frac{2}{3}(2g^j_i\delta_{(\Gamma)}{\bar P}^i_{jk;l}+g^{ij}g_{mk}\delta_{(\Gamma)} {\bar P}^m_{ij;l})
\end{eqnarray}
We now insert the explicit expressions for $\bar P^i_{jk;l}$ through the Christoffel symbols:
\begin{eqnarray}
\label{fB11}
&&=G^{l;ks}\frac{2}{3}[2g^j_i\delta_{(\Gamma)}({\bar P}^i_{jk\,,\,l}+\Gamma^i_{ls}{\bar P}^s_{jk}
-\Gamma^s_{lj}{\bar P}^i_{sk}-\Gamma^s_{lk}{\bar P}^i_{js})\nonumber\\
&&+g^{ij}g_{mk}\delta_{(\Gamma)}({\bar P}^m_{ij\,,\,l}+\Gamma^m_{ls}{\bar P}^s_{ij}-\Gamma^s_{li}
{\bar P}^m_{sj}-\Gamma^s_{lj}{\bar P}^m_{is})]\nonumber\\
&&=G^{k;l}\frac{2}{3}[2g^j_i(\delta\Gamma^i_{ls}\,{\bar P}^s_{jk}-\delta\Gamma^s_{lj}\,
{\bar P}^i_{sk}-\delta\Gamma^s_{lk}{\bar P}^i_{js})+\nonumber\\
&&\quad g^{ij}g_{mk}(\delta\Gamma^m_{ls}\,{\bar P}^s_{ij}-\delta\Gamma^s_{li}\,
{\bar P}^m_{sj}-\delta\Gamma^s_{lj}\,{\bar P}^m_{is})]
\end{eqnarray}
In the last expression we used the fact that $\delta_{\Gamma}{\bar P}^i_{jk\,,\,l}=0$
 since the term $\partial \bar P^i_{jk}/\partial x^l$ does not include $g_{ij}$.
Combining similar terms and contracting $\bar P^i_{jk}$ where it is possible back to $G_i$ we get:

\begin{eqnarray}
\label{fB12}
&&=G^{l;k}\frac{2}{3}\,(-2G_s\,\delta\Gamma^s_{lk}+g_{mk}G^s\,\delta\Gamma^m_{ls}-
2g_{mk}g^{ij}{\bar P}^m_{sj}\,\delta\Gamma^s_{li})
\end{eqnarray}

We now transition the Christoffel symbols of the second type ($\Gamma^i_{jk}$) to the Christoffel symbols 
of the first type
$\Gamma^i_{jk}=g^{im}\Gamma_{mjk}$. The variation of the term $g^{im}$ in front of $\Gamma$ could be neglected
 since the end result is proportional to $\Gamma$s and will vanish when transfered to the covariant form.

\begin{eqnarray}
\label{fB13}
&&=-\frac{4}{3}G^{k;l}G^s\delta\Gamma_{slk}+\frac{2}{3}G^{k;l}G^s\delta\Gamma_{kls}-
\frac{4}{3}G^{k;l}\ bar P^{ksi}\delta\Gamma_{sli}
\end{eqnarray}
Substituting the explicit expression for $\Gamma_{ijk}$ through $g_{ij\,,\,m}$ we get:
\begin{eqnarray}
\label{fB14}
&&=(-\frac{4}{3}G^{k;l}G^s)\frac{1}{2}(\delta g_{sk\,,\,l}+\delta g_{sl\,,\,k}-\delta g_{kl\,,\,s})\nonumber\\
&&+(\frac{2}{3}G^{k;l}G^s)\frac{1}{2}(\delta g_{kl\,,\,s}+\delta g_{ks\,,\,l}-\delta g_{ls\,,\,k})\nonumber\\
&&-(\frac{4}{3}G^{k;l}g_{mk}{\bar P}^{msi})\frac{1}{2}(\delta g_{sl\,,\,i}+\delta g_{si\,,\,l}-\delta g_{li\,,\,s})
\end{eqnarray}

The three terms of $g_{ij\,,\,l}$ in each parentheses could be written as one after renaming indices and combining similar
terms. Also note that $\bar P^{msi}$ is fully symmetrical and thus the first and the third terms in the last parentheses
cancel each other.

\begin{eqnarray}
\label{fB15}
&&=-\frac{2}{3}(G^{k;s}G^l+G^{s;l}G^k-G^{k;l}G^s)\delta g_{kl,s}\nonumber\\
&&+\frac{1}{3}(G^{k;l}G^s+G^{k;s}G^l-G^{s;l}G^k)\delta g_{kl,s}\nonumber\\
&&-(\frac{2}{3}G^{n;s}g_{mn}{\bar P}^{mkl})\frac{1}{2}\delta g_{kl\,,\,s}
\end{eqnarray}

Substituting $\bar P^{mkl}$ through vector $G_i$ we get:
\begin{eqnarray}
\label{fB16}
&&=-\frac{2}{3}(G^{k;s}G^l+G^{s;l}G^k-G^{k;l}G^s)\delta g_{kl,s}\nonumber\\
&&+\frac{1}{3}(G^{k;l}G^s+G^{k;s}G^l-G^{s;l}G^k)\delta g_{kl,s}\nonumber\\
&&-(\frac{2}{3}G^{n;s}g_{mn})\frac{1}{6}(\underline{G^mg^{kl}}+G^kg^{ml}+G^lg^{km})\delta g_{kl,s}
\end{eqnarray}

The underlined term vanishes due to the constraint $G^kG_k=1$ - $G^{k;l}G_k=1/2(G^kG_k)^{;l}=0$.
We now apply the "partial integration rule" of the variational method and transfer $\partial_s$ from the metric
tensor to the term in front of it with the opposite sign:

\begin{eqnarray}
\label{fB17}
&&=-[-\frac{2}{3}(G^{k;s}G^l+G^{s;l}G^k-G^{k;l}G^s)]_{\,,\,s}\,\delta g_{kl}\nonumber\\
&&-[+\frac{1}{3}(G^{k;l}G^s+G^{k;s}G^l-G^{s;l}G^k)]_{\,,\,s}\,\delta g_{kl}\nonumber\\
&&-[-(\frac{2}{3}G^{n;s}g_{mn})\frac{1}{6}(G^kg^{ml}+G^lg^{km})_{\,,\,s}]\,\delta g_{kl}
\end{eqnarray}

We can now replace the partial derivatives with the covariant ones. Tensor $\delta g_{kl}$ is symmetrical and so should be
the expression in the front of it. After symmetrization we get:

\begin{eqnarray}
\label{fB18}
&&=\{\frac{1}{3}[(G^{s;k}G^l+G^{s;l}G^k)+(G^{k;s}G^l+G^{l;s}G^k)-(G^{k;l}G^s+G^{l;k}G^s)\nonumber\\
&&+\frac{1}{6}[(G^{s;k}G^l-G^{s;l}G^k)+(G^{k;s}G^l+G^{l;s}G^k)-(G^{k;l}G^s+G^{l;k}G^s)]\nonumber\\
&&+\frac{1}{9}[(G^{k;s}G^l+G^{l;s}G^k)]\}_{;s}\delta g_{kl}
\end{eqnarray}
And collecting similar terms we get:

\begin{eqnarray}
\label{fB18a}
&&=[\frac{1}{2}(G^{s;k}G^l+G^{s;l}G^k)+\frac{5}{18}(G^{k;s}G^l+G^{l;s}G^k)\nonumber\\
&&-\frac{1}{2}(G^{k;l}G^s+G^{l;k}G^s)]_{;s}\delta g_{kl}
\end{eqnarray}
From which follows the following expression for $q^{kl}_{(\Gamma)}$

\begin{eqnarray}
\label{fB19}
&&q_{\Gamma}^{kl}:=\delta_{(\Gamma)} L_2/\delta g_{kl}=\nonumber\\
&&=[\frac{1}{2}(G^{s;k}G^l+G^{s;l}G^k)+\frac{5}{18}(G^{k;s}G^l+G^{l;s}G^k)\nonumber\\
&&\quad \quad -\frac{1}{2}(G^{k;l}G^s+G^{l;k}G^s)]_{;s}
\end{eqnarray}

Combining the expression for $q_{g}^{kl}$, eq. (\ref{fB9}), and the term $q_{\Gamma}^{kl}$, eq. (\ref{fB19}),
 we get the final result for the
variation of the Lagrangian $L_2=G^{k;l}G_{k;l}$ with respect to $g_{kl}$:
\begin{eqnarray}
\label{fB20}
&&q^{kl}=\frac{\partial (L_2\sqrt{g})}{\sqrt{g}\partial g_{kl}}\nonumber\\
&&=-\frac{5}{9}G^{k;m}G^{l;n}g_{mn}-G^{m;k}G^{n;l}g_{mn}+\frac{7}{18}G_{m;n}G^{m;n}g^{kl}+\nonumber\\
&&=[\frac{1}{2}(G^{s;k}G^l+G^{s;l}G^k)+\frac{5}{18}(G^{k;s}G^l+G^{l;s}G^k)\nonumber\\
&&-\frac{1}{2}(G^{k;l}G^s+G^{l;k}G^s)]_{;s}
\end{eqnarray}

\newpage
{\it \underline{Appendix C:}}
\vskip 0.5em

The goal of this appendix is to derive the equation $q_{kl}=0$ , where $q_{kl}$
 is a variation of the Lagrangian $L_{(G)}$ with respect to $g_{kl}$. In this appendix we will do the calculations for 
the Lagrangian $L_3=(G^m_{;m})^2$.

We begin by writing the Lagrangian as a function of $P^i_{jk}$:
\begin{eqnarray}
\label{fC1}
&&G_k=\frac{1}{3}(2{\bar P}^i_{jk}g^j_i+{\bar P}^m_{st}g_{mk}g^{st})\\
\label{fC1a}
&&and\quad G_{k;l}=\frac{1}{3}(2{\bar P}^i_{jk;l}g^j_i+{\bar P}^m_{st;l}g_{mk}g^{st})\\
\label{fC1b}
&&{\bar P}_{ijk}=\frac{1}{6}(G_ig_{jk}+G_jg_{ik}+G_kg_{ij})r\\
\label{fC1c}
&&L_3=(G^m_{;m})^2
\end{eqnarray}
Here $\bar P^i_{jk}$ is the symmetric part in low indices of the $P^i_{jk}$ tensor and
 $g^i_j$ are Kronneker symbols ($g^i_j=1$, if $i=j$ and $g^i_j=0$ otherwise).

The variation of $L_3$ with respect to $g_{kl}$ consists of two parts. The first part is with respect to algebraic terms
 (including $\sqrt{g}$), which we denote as $q_{(g)}^{kl}$. The second part is with respect to the partial derivative
 of $g_{ij}$ ($g_{ij,k}$)
 that are part of Christoffel symbols  ${\Gamma}^i_{jk}$, which we denote as $q_{(\Gamma)}^{kl}$.

The $q_{(g)}^{kl}$ is a straight partial derivative of $L_3$ by $g_{kl}$:
\begin{eqnarray}
\label{fC2}
&&q_{(g)}^{kl}:=\frac{1}{\sqrt{g}} \partial (\sqrt{g}{G^m_{;m}}^2)/\partial g_{kl}=
2{G^m_{;m}} \partial {G^m_{;m}}/\partial g_{kl} + \frac{1}{2} {G^m_{;m}}^2 g^{kl} \nonumber\\
&&=\frac{1}{2} {G^m_{;m}}^2 g^{kl}+\frac{2}{3}{G^m_{;m}} \partial (2{\bar P}^i_{jk;l}g^j_ig^{kl}
+{\bar P}^m_{st;l}g_{mk}g^{st}g^{kl})/\partial g_{kl}\\
&&=\frac{1}{2} {G^m_{;m}}^2 g^{kl}+\frac{2}{3}{G^m_{;m}} \partial (2{\bar P}^i_{jk;l}g^j_ig^{kl}
+{\bar P}^m_{kl;m}g^{kl})/\partial g_{kl}\nonumber\\
&&=\frac{1}{2} {G^m_{;m}}^2 g^{kl}-\frac{4}{3}{G^m_{;m}} {\bar P}^i_{jm;n}g^j_ig^{mk}g^{nl}
-\frac{2}{3} {G^m_{;m}} {\bar P}^s_{mn;s}g^{mk}g^{nl}\nonumber
\end{eqnarray}

Substituting $\bar P^i_{jk}$ through $G_i$, eq. (\ref{fC1b}), in the expression above we get:

\begin{eqnarray}
\label{fC3}
&&q_{(g)}^{\,kl}=\frac{1}{2} {G^m_{;m}}^2 g^{kl}-\frac{4}{3}{G^m_{;m}} G^{k;l} \nonumber\\
&&-\frac{1}{9} {G^m_{;m}} [(G^{s}g^{kl})_{;s}+(G^{k}g^{sl})_{;s}+(G^{l}g^{ks})_{;s}] \\
&&=[(\frac{1}{2}-\frac{1}{9}){G^m_{;m}}^2 g^{kl}+(-\frac{2}{3}-\frac{1}{9})G^m_{;m}(G^{k;l}+G^{l;k})]\nonumber
\end{eqnarray}

From which it follows the final expression for the $q_{(g)}^{kl}$:
\begin{eqnarray}
\label{fC4}
&&q_{(g)}^{kl}=\frac{7}{18}{G^m_{;m}}^2 g^{kl}-\frac{7}{9}G^m_{;m}(G^{k;l}+G^{l;k})
\end{eqnarray}
As a check point, the contraction of $q^{(g)kl}$ should be zero - as it is. 
It is due to the quadratic form of the Lagrangian $L_3$ with respect to the metric tensor $g^{ij}$

\vskip 3em
We now derive the expression for the variation of the Lagrangian $L_3$ ($L_3={G^k_{;k}}^2$) with respect to partial 
derivatives of $g_{ij}$ ($g_{ij,k}$), which are part of the Christoffel symbols  ${\Gamma}^i_{jk}$ - denoted
 as $q_{(\Gamma)}^{kl}$. 
The simplest way of doing this is to calculate the deviation $\delta_{(\Gamma)} L_3$. 
Again, as above, we start with writing $L_3$ as a function of $P^i_{jk}$.

\begin{eqnarray}
\label{fC5}
&&\delta q_{(\Gamma)}^{kl}=\delta {G^m_{;m}}^2=2{G^m_{;m}} \delta {G^m_{;m}}=
\frac{2}{3}{G^m_{;m}} \delta [g^{kl} (2\bar P^i_{jk;l}g^j_i+\bar P^n_{ij;l}g^{ij}g_{nk})]\nonumber\\
&&=\frac{4}{3} {G^m_{;m}} g^{kl}g^j_i \delta \bar P^i_{jk;l}+\frac{2}{3}{G^m_{;m}} g_{nk}g^{ij}g^{kl}
 \delta \bar P^n_{ij;l}\nonumber\\
&&=\frac{4}{3}{G^m_{;m}} g^{kl}g^j_i \delta(\bar P^i_{jk,l}+\Gamma ^i_{ls}\bar P^s_{jk}-\Gamma ^s_{lj}\bar P^i_{sk}-
\Gamma ^s_{lk}\bar P^i_{js})\nonumber\\
&&+\frac{2}{3}{G^m_{;m}} g^l_n g^{ji} \delta(\bar P^n_{ij,l}+\Gamma ^n_{ls}\bar P^s_{ij}-\Gamma ^s_{li}\bar P^n_{sj}-
\Gamma ^s_{lj}\bar P^n_{is})\nonumber\\
&&=\frac{4}{3}{G^m_{;m}} g^{kl}g^j_i (\bar P^s_{jk}\delta \Gamma ^i_{ls}-\bar P^i_{sk}\delta \Gamma ^s_{lj}-
\bar P^i_{js}\delta \Gamma ^s_{lk})\nonumber\\
&&+\frac{2}{3}{G^m_{;m}} g^l_n g^{ji}(\bar P^s_{ij}\delta \Gamma ^n_{ls}-\bar P^n_{sj}\delta \Gamma ^s_{li}-
\bar P^n_{is}\delta \Gamma ^s_{lj})
\end{eqnarray}

In the expression above the term $\delta \bar P^i_{jk,l}$ vanishes since it does not contains the metric tensor.
The expression above can be simplfied, if we reduce, where it possible, the indices and remembering that
 contraction of $\bar P^i_{jk}$ in any two indices is $G_i$. 

We will also replace Christoffel  symbols of 
the second type ($\Gamma ^i_{jk}$) with Christoffel  symbols of the first type 
$\Gamma^i_{jk}=g^{im}\Gamma_{mjk}$. The variation of the term $g^{im}$ in front of $\Gamma$ could be neglected
 since the end result is proportional to $\Gamma$s and will vanish when transfered to the covariant form.
After having done this we get:
\begin{eqnarray}
\label{fC6}
&&=-\frac{4}{3}{G^m_{;m}} g^{kl} G^s \delta \Gamma _{skl}+\frac{2}{3}{G^m_{;m}} g^{kl}G^s \delta \Gamma _{kls}
-\frac{4}{3}{G^m_{;m}} \bar P^{kls} \delta \Gamma _{kls}
\end{eqnarray}

Substituting in eq. (\ref{fC6}) expression for Christoffel  symbols through partial derivatives 
$\Gamma _{ijk}=1/2(g_{ij,k}+g_{ik,j}-g_{jk,i})$ we get:
\begin{eqnarray}
\label{fC7}
&&=-\frac{2}{3}{G^m_{;m}} g^{kl} G^s (\delta g_{sl,k}+\delta g_{sk,l}-\delta  g_{kl,s})\nonumber\\
&&+\frac{1}{3}{G^m_{;m}} g^{kl} G^s (\delta g_{kl,s}+\underline{\delta g_{ks,l}-\delta  g_{sl,k}})\\
&&-\frac{2}{3}{G^m_{;m}} \bar P^{kls} (\delta g_{kl,s}+\underline{\delta g_{ks,l}-\delta  g_{sl,k}})\nonumber
\end{eqnarray}

The underlined terms cancel each other out. Replacing $\bar P^{kls}$ with its representation through $G_i$, eq. (\ref{fC1b}) 
and using the rule of partial integration we can transfer the partial derivative from the metric tensor to the
term in font of it and rewrite the above expression in this manner:
\begin{eqnarray}
\label{fC7a}
&&=\frac{2}{3}({G^m_{;m}} g^{kl} G^s)_{,k} \delta g_{sl}+\frac{2}{3}({G^m_{;m}} g^{kl})_{,l}\delta g_{sk}-
\frac{2}{3}({G^m_{;m}} g^{kl})_{,s}\delta  g_{kl}\nonumber\\
&&+\frac{1}{3}({G^m_{;m}} g^{kl} G^s)_{,s} \delta g_{kl}\\
&&-\frac{1}{9}({G^m_{;m}} (G^kg^{ls}+G^lg^{ks}+G^sg^{kl})_{,s} \delta g_{kl}\nonumber\\
&&=\{(\frac{2}{3}+\frac{1}{9})[({G^m_{;m}} G^k)^{;l}+({G^m_{;m}} G^l)^{;k}]+(-\frac{2}{3}-\frac{1}{3}+
\frac{1}{9})({G^m_{;m}} G^sg^{kl})_{;s} \} \delta g_{kl}\nonumber\\
&&=[\frac{7}{9}[({G^m_{;m}} G^k)^{;l}+({G^m_{;m}} G^l)^{;k}]-\frac{8}{9}({G^m_{;m}} G^sg^{kl})_{;s}]\delta g_{kl}\nonumber
\end{eqnarray}

From which it follows that the variation of Lagrangian $L_3$ with respect to $g_{kl}$ present in the terms containing
Christoffel symbols  ($q_{(\Gamma)}^{kl}$) has this form:
\begin{eqnarray}
\label{fC8}
&&q_{(\Gamma)}^{kl}=\frac{7}{9}[({G^m_{;m}} G^k)^{;l}+({G^m_{;m}} G^l)^{;k}]-\frac{8}{9}({G^m_{;m}} G^sg^{kl})_{;s}
\end{eqnarray}

And finally, combining eq. (\ref{fC4}) and eq.(\ref{fC8}) we get the final expression for the variation of the
Lagrangian $L_3$ with respect to the metric tensor $g_{kl}$:
\begin{eqnarray}
\label{fC8a}
&&q^{kl}=q_{(g)}^{kl}+q_{(\Gamma)}^{kl}\nonumber\\
&&=\frac{7}{18}{G^m_{;m}}^2 g^{kl}-\frac{7}{9}G^m_{;m}(G^{k;l}+G^{l;k})\nonumber\\
&&+\frac{7}{9}[({G^m_{;m}} G^k)^{;l}+({G^m_{;m}} G^l)^{;k}]-\frac{8}{9}({G^m_{;m}} G^sg^{kl})_{;s}
\end{eqnarray}

\newpage
{\it \underline{Appendix D:}}
\vskip 1em
In this appendix we will derive the expression for the main quantities, eq. (\ref{f37}) through eq. (\ref{f37c}) 
for the sphere symmetrical problem. We will use new variables as indicated by (\ref{f36})
\begin{eqnarray}
\label{fD1}
&&a)\quad \hat {G_1}=\frac{G_1^2 g_0r}{g_1}\nonumber\\
&&b)\quad x=1/r\\
&&c)\quad \hat g=g_0g_1\nonumber\\
&&d) \,\, the  \,\, derivative \,\,  by \,\, x \,\, is \,\, designated \,\, as \,\, (\,')\nonumber
\end{eqnarray}

Often during calculations it is convenient to use normalized $\bar G_0$ and $\bar G_1$ variables 
according to these formulas:
\begin{eqnarray}
\label{fD2}
&&\bar G_0=G_0/\sqrt{g_0} \quad \bar G_1=G_1/\sqrt{g_1}
\end{eqnarray}

The requirement $G^iG_i=1$ will have this form:
\begin{eqnarray}
\label{fD3}
&&(\bar G_0)^2-(\bar G_1)^2=1
\end{eqnarray}

And after differentiating by r we will have:
\begin{eqnarray}
\label{fD4}
&&\bar G_{0,1}\bar G_0=\bar G_{1,1}\bar G_1
\end{eqnarray}

We also notice that the 2-index tensor $G_{i;j}$ has six non zero components: $G_{0;0}$, $G_{1;0}$, $G_{0;1}$, $G_{1;1}$,
 $G_{2;2}$ and $G_{3;3}=G_{2;2}sin(\theta)$
\vskip 1em
For the Lagrangian $L_{12}=L_1-L_2$ we have:
\begin{eqnarray}
\label{fD5}
&&L_{12}:=G_{m;n}G^{m;n}-G_{m;n}G^{n;m}=\nonumber\\
&&[{G_{00}}^2(g^{00})^2+{G_{11}}^2(g^{11})^2+{G_{01}}^2g^{00}g^{11}+{G_{10}}^2g^{00}g^{11}+2{G_{22}}^2(g^{22})^2]\nonumber\\
&&-[{G_{00}}^2(g^{00})^2+{G_{11}}^2(g^{11})^2+2G_{01}G_{10}g^{00}g^{11}+2{G_{22}}^2(g^{22})^2]\nonumber\\
&&=(G_{01}-G_{10})^2g^{00}g^{11}=-\frac{1}{g_0g_1} (G_{0,1})^2=-\frac{x^4}{\hat g}(G_0\,')^2
\end{eqnarray}
And thus the final resul for $L_{12}$ is:
\begin{eqnarray}
\label{fD6}
&&L_{12}:=G_{m;n}G^{m;n}-G_{m;n}G^{n;m}=-\frac{x^4}{\hat g}(G_0\,')^2
\end{eqnarray}

For the Lagrangian $L_{32}=L_3-L_2$ we have:
\begin{eqnarray}
\label{fD7}
&&L_{32}:=L_3-L_2=(G^m_{;m})^2-G_{m;n}G^{n;m}\nonumber\\
&&=[(G_{00}g^{00}+G_{11}g^11+2G_{22}g^{22}]^2\nonumber\\
&&\quad -[{G_{00}}^2(g^{00})^2+{G_{11}}^2(g^{11})^2+2G_{01}G_{10}g^{00}g^{11}+2{G_{22}}^2(g^{22})^2]\nonumber\\
&&=2(G_{22}g^{22})^2+2G_{00}G_{11}g^{00}g^{11}+4G_{00}G_{22}g^{00}g^{22}\nonumber\\
&&\quad +4G_{11}G_{22}g^{11}g^{22}-2G_{01}G_{10}g^{00}g^{11}\nonumber\\
&&=2(-\Gamma^1_{22}g^{22}G_1)^2+2(-\Gamma^1_{00}G_1)(G_1-\Gamma^1_{11}G_1)g^{00}g^{11}\nonumber\\
&&\quad +4(\Gamma^1_{00}G_1)(\Gamma^1_{22}G_1)+4(G_{1,1}-\Gamma^1_{11}G_1)(-\Gamma^1_{22}G_1)g^{11}g^{22}\nonumber\\
&&\quad -2(G_{0,1}-\Gamma^0_{01}G_0)(-\Gamma^0_{01}G_0)g^{00}g^{11}\nonumber\\
&&=\frac{1}{2}(\frac{g_{2,1}}{g_2})^2\frac{1}{g_1^2}G_1^2+
(\frac{g_{0,1}}{g_0})(\frac{G_1}{\sqrt{g_1}})_{,1}G_1\sqrt{g_1}\frac{1}{g_1^2}
+(\frac{g_{0,1}}{g_0})(\frac{g_{2,1}}{g_2})G_1^2 \frac{1}{g_1^2}\nonumber\\
&&\quad +2(\frac{g_{2,1}}{g_2})(\frac{G_1}{\sqrt{g_1}})_{,1}G_1\sqrt{g_1}\frac{1}{g_1^2}
+(\frac{g_{0,1}}{g_0})(\frac{G_0}{\sqrt{g_0}})_{,1}G_0\sqrt{g_0}(-\frac{1}{g_1})\nonumber\\
&&=\frac{1}{g_1}[\frac{1}{2} (\frac{g_{2,1}}{g_2})^2 \bar G_1^2+\underline{\bar G_{1,1} G_1(\frac{g_{0,1}}{g_0})}
+(\frac{g_{2,1}}{g_2})(\frac{g_{0,1}}{g_0}) \bar G_1^2\nonumber\\
&&\quad +2\bar G_{1,1} G_1(\frac{g_{2,1}}{g_2})-\underline{\bar G_{0,1}G_0 (\frac{g_{0,1}}{g_0})}]\nonumber\\
&&=\frac{1}{g_1}(\frac{g_{2,1}}{g_2})[(\bar G_1^2)_{,1}+\frac{g_{0,1}}{g_0}G_1^2+
\frac{1}{2}\frac{g_{2,1}}{g_2}G_1^2]\nonumber\\
&&=\frac{1}{g_1}[\frac{(\bar G_1^2\sqrt{g_2} g_0)_{,1}}{\sqrt{g_2} g_0}]\frac{g_{2,1}}{g_2}=
-\frac{2 (\bar G_1^2rg_0)^{\,'}}{r^4g_0g_1}
\end{eqnarray}
where the underlined terms cancel each other.

And thus the final result for $L_{32}$ is:
\begin{eqnarray}
\label{fD8}
&&L_{32}:=(G^m_{;m})^2-G_{m;n}G^{n;m}=-\frac{2x^4 (\hat G_1)^{\,'}}{\hat g}
\end{eqnarray}

\vskip 1em
We now derive the expression for $J=(G^{m;n}G_n)_{;m}$. It is in the form of a divergence of a vector, which also
can be written as:
\begin{eqnarray}
\label{fD9}
&&J=(G^{1;n}G_n\sqrt{g})_{,1}/\sqrt{g}=(G_{1;n}G^ng^{11}\sqrt{g})_{,1}/\sqrt{g}\\
&&\,\, where \,\, g={g_0g_1}\,g_2\nonumber
\end{eqnarray}
The term $G_{1;n}G^n$ can be rearranged as the following:
\begin{eqnarray}
\label{fD10}
&&G_{1;n}G^n=(G_{1;0}G_0g^{00}+G_{1;1}G_1g^{11})g^{11}\nonumber\\
&&=(-\Gamma^0_{01})G_)g^{00}+(G_{1,1}-\Gamma^1_{11})G_1g^{11}\nonumber\\
&&=-\frac{1}{2} \frac{g_{0,1}}{g_0} \frac{G_0^2}{g_0}+(\frac{G_1}{\sqrt{g_1}})_{,1}\sqrt{g_1}G_1(-\frac{1}{g_1})\nonumber\\
&&=-\frac{1}{2} \frac{g_{0,1}}{g_0}\bar G_0^2-\bar G_{1,1}\bar G_1=
-\frac{1}{2} \frac{g_{0,1}}{g_0}\bar G_0^2-\bar G_{0,1}\bar G_0\nonumber\\
&&=-\frac{1}{2}(\bar G_0^2g_0)_{,1}\frac{1}{g_0}=-\frac{1}{2}(G_0^2)_{,1}\frac{1}{g_0}
\end{eqnarray}
Substituting eq. (\ref{fD10}) in eq. (\ref{fD9}) we will get:
\begin{eqnarray}
\label{fD11}
&&J=[-\frac{1}{2}(G_0^2)_{,1}\frac{1}{g_0}(-\frac{1}{g_1})\sqrt{g_0g_1}\,g_2]_{,1} \frac{1}{\sqrt{g_0g_1}\,g_2}
\end{eqnarray}

And switching to variable "x" we get the final result for $J$:
\begin{eqnarray}
\label{fD12}
&&J=\frac{x^4}{2\sqrt{\hat g}}[(G_0^2)^{\,'}\frac{1}{\sqrt{\hat g}}]^{\,'}
\end{eqnarray}

\vskip 1em
We now derive the expression for $\bar J=(G^m_{;m}G^n)_{;n}$. It is in the form of a divergence of a vector, which also
can be written as:
\begin{eqnarray}
\label{fD12a}
&&\bar J=(G^m_{;m}G_1g^{11}\sqrt{g})\frac{1}{\sqrt{g}}\\
&&\,\, where \,\, g={g_0g_1}\,g_2\nonumber
\end{eqnarray}

The expression for $G^m_{;m}$ in terms of $\bar G_1$ has this form:
\begin{eqnarray}
\label{fD13}
&&G^m_{;m}=(G_1g^{11}\sqrt{g})\frac{1}{\sqrt{g}}\nonumber\\
&&\,\, where \,\, g=\sqrt{g_0g_1}\,g_2\nonumber\\
&&=-(\bar G_1 \sqrt{g_0}\,g_2)_{,1}\frac{1}{\sqrt{g_1g_2}\,g_2}\nonumber\\
&&=-\frac{1}{\sqrt{g_1}}(\bar G_{1,1}+\frac{g_{0,1}}{2g_0}\bar G_1+\frac{g_{2,1}}{g_2}\bar G_1)
\end{eqnarray}

Substituting this expression in eq. (\ref{fD12a}) we will get:
\begin{eqnarray}
\label{fD14}
&&\bar J=\frac{1}{\sqrt{g_1g_0}\,g_2}[\frac{1}{\sqrt{g_1}}(\bar G_{1,1}+\frac{g_{0,1}}{2g_0}G_1+\frac{g_{2,1}}{g_2}G_1)
\bar G_1 \sqrt{g_1} \frac{\sqrt{g_1g_0}\,g_2}{g_1}]_{,1}\nonumber\\
&&\quad =\frac{1}{\sqrt{g_1g_0}\,g_2}[(\frac{1}{2} \bar G_1^2+\frac{g_{0,1}}{2g_0}\bar G_1^2+\frac{g_{2,1}}{g_2}\bar G_1^2) 
\frac{\sqrt{g_0}\,g_2}{\sqrt{g_1}}]_{,1}\nonumber\\
&&\quad =\frac{1}{2\sqrt{g_1g_0}\,g_2}[(\bar G_1^2 g_0g_2^2)_{,1}\frac{1}{g_0g_2^2}
\frac{\sqrt{g_0}\,g_2}{\sqrt{g_1}}]_{,1}\nonumber\\
&&\quad =\frac{1}{2\sqrt{\hat g}\,g_2}[(\hat G_1 g_2^{3/2})_{,1}\frac{1}{\sqrt{\hat g}\,g_2}]_{,1}\nonumber\\
&&\quad =[\frac{1}{2\sqrt{\hat g}\,g_2}(\hat G_{1,1} \sqrt{ g_2}+\frac{3}{2} \frac{g_{2,1}}
{\sqrt{g_2}} \hat G)\frac{1}{\sqrt{\hat g}}]_{,1}
\end{eqnarray}
And replacing $g_2$ with $r^2$ and switching to variable x (x=1/r) we get:

\begin{eqnarray}
\label{fD15}
&&\bar J=\frac{1}{2\sqrt{\hat g}\,r^2}(\hat G_{1,1} r+3\hat G)\frac{1}{\sqrt{\hat g}}]=
\frac{x^4}{2\sqrt{\hat g}} [(\hat G_1\,' x-3\hat G)\frac{1}{\sqrt{\hat g}}]\,'
\end{eqnarray}
And thus the final expression for $\bar J$ is:

\begin{eqnarray}
\label{fD16}
&&\bar J=(G^m_{;m}G^n)_{;n}=
\frac{x^4}{2\sqrt{\hat g}} [(\hat G_1\,' x-3\hat G)\frac{1}{\sqrt{\hat g}}]\,'
\end{eqnarray}

\newpage
{\it \underline{Appendix E:}}
\vskip 1em
Here we will derive the expression for the equation $q_{22}=0$.
We begin with eq. (\ref{f25}) 

\begin{eqnarray}
\label{fE1}
&&q^{kl}:=\frac{1}{\sqrt{g}}\delta (\sqrt{g}L_G)/ \delta{g_{kl}}+T^{kl}=0\quad,\,\,where\nonumber\\
&&q_{kl}=\lambda_1[\frac7{18}(G^{m;n}G_{m;n})g_{kl}-{G^s}_{;k}G_{s;l}-\frac{5}{9}{G_k}^{;s}G_{l;s}]+\nonumber\\
&&\quad \,\,+\lambda_2[\frac7{18}(G^{m;n}G_{n;m})g_{kl}-\frac79 G_{k;s}{G^s}_{;l}-\frac79 G_{l;s}{G^s}_{;k}]\nonumber\\
&&\quad \,\, +\lambda_3[\frac7{18}({(G^m{;m})}^2g_{kl}-\frac79 G^s_s G_{k;l}-\frac79 G^s_s G_{l;k}]\nonumber\\
&&\quad \,\,+\lambda_1[\frac12{(G_{s;k}G_l+G_{s;l}G_k)}^{;s}+\frac5{18}{(G_{k;s}G_l+G_{l;s}G_k)}^{;s}\nonumber\\
&&\quad \,\, -\frac12 {(G_{k;l}G_s+G_{l;k}G_s)}^{;s}]\nonumber\\
&&\quad \,\,+\lambda_2[\frac19{(G_{m;n}G^n)}^{;m}g_{kl}+\frac5{18}{(G_{s;k}G_l+G_{s;l}G_k)}^{;s}\nonumber\\
&&\quad \,\,+\frac12{(G_{k;s}G_l+G_{l;s}G_k)}^{;s}-\frac12 {(G_{k;l}G_s+G_{l;k}G_s)}^{;s}]\nonumber\\
&&\quad \,\,+\lambda_3[-\frac89{(G^m_{;m}G^s)}_{;s}g_{kl}+\frac79{(G^m_{;m}G_k)}_{;l}+\frac79{(G^m_{;m}G_l)}_{;k}]\nonumber\\
&&\quad \,\,+\frac19T(g_{kl}+5G_kG_l)]
\end{eqnarray}

In the last three square brackets we will open the parentheses and take the covariant derivatives of each term.
Taking into account that $G_2=G_3=0$ the expression for each term can be written as:
\begin{eqnarray}
\label{fE2}
&&{(G_{s;k}G_l+G_{s;l}G_k)}^{;s}=\nonumber\\
&&\quad ({G_{s;k}}^{;s}G_l+{G_{s;l}}^{;s}G_k)+G_{s;k}{G_l}^{;s}+G_{s;l}{G_k}^{;s}=2G_{s;2}{G_2}^{;s}
\end{eqnarray}
\begin{eqnarray}
\label{fE2a}
&&{(G_{k;s}G_l+G_{l;s}G_k)}^{;s}=\nonumber\\
&&\quad ({G_{k;s}}^{;s}G_l+{G_{l;s}}^{;s}G_k)+G_{k;s}{G_l}^{;s}+G_{l;s}{G_k}^{;s}=2G_{2;s}{G_2}^{;s}
\end{eqnarray}
\begin{eqnarray}
\label{fE2b}
&&{(G_{k;l}G_s+G_{l;k}G_s)}^{;s}=\nonumber\\
&&\quad (G_{k;l;s}G^{s}+G_{l;k;s}G^s)+G_{k;l}{G_s}^{;s}+G_{l;k}{G_s}^{;s}=\nonumber\\
&&\quad 2G_{2;2;s}G^s+2G_{2;2}{G_s}^{;s}\\
&&\nonumber\\
\label{fE2c}
&&{(G^m_{;m}G_k)}_{;l}=G^m_{;m;l}G_k+G^m_{;m}G_{k;l}=2G^m_{;m}G_{2;2}
\end{eqnarray}

Substituting eq. (\ref{fE2}) through (\ref{fE2c}) into eq.(\ref{fE1}) we get:
\begin{eqnarray}
\label{fE3a}
&&g_{22}[\lambda_1 \frac{7}{18} L_1+\lambda_2 \frac{7}{18} L_2+\lambda_3 \frac{7}{18} L_3+
\frac{\lambda_2J}{9}-\frac{\lambda_38\bar J}{9}+\frac{T}{9}]\\
&&G_{2;2}g^{22}[-\lambda_1-\frac59 \lambda_1-\frac{14}{9} \lambda_2+\lambda_1+\frac{10}{18}\lambda_1
+\frac{10}{18}\lambda_2+\lambda_2]\\
&&+G^m_{;m}G_{2;2}[-\frac{14}\lambda_3-\lambda_1-\lambda_2+\frac{14}{9}\lambda_3]+
(-\lambda_1-\lambda_2)G_{2;2;s}G^s
\end{eqnarray}
where $L_1$, $L_2$, $L_3$ are the three terms of the Lagrangian and $J$, $\bar J$ are the scalars defined below:
\begin{eqnarray}
\label{fE4}
&&L_G=\lambda_1L_1+\lambda_2L_2+\lambda_3L_3 \nonumber\\
&&L_1=G_{k;l}G^{k;l}\quad L_2=G_{k;l}G^{l;k}\quad L_3=(G^m_{;m})^2\\
&&J={(G_{k;l}G^l)}^k\quad \bar J={(G^m_{;m}G^l)}_{;l}
\end{eqnarray}

Substituting $T$ for its expression through $L_G$, $J$, and $\bar J$, eq. (\ref{f28}), we get:
\begin{eqnarray}
\label{fE5}
&&q_{22}=g_{22}(\frac{1}{2} L_G-\lambda_3\bar J)+(-\lambda_1-\lambda_2)(G_{2;2;s}G^s+G^m_{;m}G_{2;2})
\end{eqnarray}

With the assumption that $\lambda_2=-\lambda_1-\lambda_3$ and using the formula $L_G=-\lambda_1J+2\lambda_3\bar J$ -
eq. (\ref{f27}), (\ref{f28}) - we get:
\begin{eqnarray}
\label{fE6}
&&q_{22}=-\frac{\lambda_1 J}{2} g_{22}+\lambda_3(G_{2;2;s}G^s+G^m_{;m}G_{2;2})=0
\end{eqnarray}

We now calculate the explicit expressions for the terms in eq. (\ref{fE6}) above. 
For the first term in the parentheses we have:
\begin{eqnarray}
&&G_{2;2;s}G^s=G_{2;2;0}G^0+G_{2;2;1}G_1g^{11}=G_{2;2;1}G_1g^{11}
\end{eqnarray}
because
\begin{eqnarray}
\label{fE7c}
&&G_{2;2;0}={(G_{2;2})}_{,0}-\Gamma ^s_{02} G_{s;2}-\Gamma ^s_{02} G_{2;s}=0
\end{eqnarray}
Thus
\begin{eqnarray}
\label{fE7d}
&&G_{2;2;s}G^s=G_{2;2;1}G_1g^{11}=\nonumber\\
&&=[{(G_{2;2})}_{,1}-{\Gamma}^s_{12} G_{s;2}-{\Gamma}^s_{12} G_{2;s}]G_1g^{11}=\nonumber\\
&&\quad \{s=2\}=[{(G_{2;2})}_{,1}-2{\Gamma}^2_{12} G_{2;2}]G_1g^{11}\nonumber\\
&&=[{(G_{2;2})}_{,1}-\frac{g_{22,1}}{g_{22}} G_{2;2}]G_1g^{11}=g_{22}{(\frac{G_{_{2;2}}}{g_{22}})}_{,1}G_1g^{11}\nonumber\\
&&=\frac{g_{22}}{g_1^2}[G_1G_{1,1}\frac{g_{2,1}}{2g_2}-G_1^2\frac{g_{1,1}}{2g_1}\frac{g_{2,1}}{2g_2}
+G_1^2{(\frac{g_{2,1}}{2g_2})}_{,1}]
\end{eqnarray}

For the second term in the parentheses of eq. (\ref{fE6}) we have:
\begin{eqnarray}
\label{fE7}
&&G^k_{;k}=\frac{ {(G_1g^{11}g_2\sqrt{g_0g_1})}_{,1} }{g_2\sqrt{g_0g_1}}\nonumber\\
&&=-\frac{1}{g_1}[G_{1,1}+\frac{g_{0,1}}{2g_0}G_1-\frac{g_{1,1}}{2g_1}G_1+\frac{g_{2,1}}{g_2}G_1]
\end{eqnarray}
\begin{eqnarray}
\label{fE7a}
&&G_{2;2}=-{\Gamma}^1_{22} G_1=\frac{1}{2} g_{22}g^{11}G_1=-(\frac{1}{g_1} \frac{g_{2,1}}{2g_2} G_1)g_{22}
\end{eqnarray}
\begin{eqnarray}
\label{fE7b}
&&G^k_{;k}G_{22}=\frac{g_{22}}{g_1^2}[G_1G_{1,1}+\frac{g_{0,1}}{2g_0}G_1^2+\frac{g_{2,1}}{g_2}G_1^2]\frac{g_{2,1}}{2g_2}
\end{eqnarray}

Combining eq.(\ref{fE7d}) and (\ref{fE7b}) we get:
\begin{eqnarray}
\label{fE8}
&&(G_{2;2;s}G^s+G^m_{;m}G_{2;2})\frac{1}{g_{22}}=\nonumber\\
&&\frac{1}{g_1^2}[G_1G_{1,1}+\frac{g_{0,1}}{4g_0}G_1^2-\frac{3g_{1,1}}{4g_1}G_1^2]\frac{g_{2,1}}{g_2}\nonumber\\
&&+\frac{1}{2}{(\frac{g_{2,1}}{g_2})}^2G_1^2+{(\frac{g_{2,1}}{2g_0})}_{,1}G_1^2\nonumber\\
&&=\frac{1}{g_1^2}\frac{g_{2,1}}{g_2}[\frac{1}{2}{(G^2)}_{,1}+\frac{g_{0,1}}{4g_0}G_1^2-\frac{3g_{1,1}}{2g_1}G_1^2+
\frac{g_{2,1}}{4g_2}]\nonumber\\
&&=\frac{g_{2,1}}{2g_2g_1^2}{(\frac{G_1^2\sqrt{g_0}\sqrt{g_2}}{\sqrt{g_1^3}})}_{,1}
\frac{\sqrt{g_1^3}}{\sqrt{g_0}\sqrt{g_2}}\nonumber\\
&&=\frac{1}{r^2\sqrt{\hat g}}{(\frac{\hat G_1g_0r}{g_1\sqrt{\hat g}})}_{,1}=
-\frac{1}{r^4\sqrt{\hat g}}{(\frac{\hat G_1}{\sqrt{\hat g}})}'
\end{eqnarray}
And thus the final result for equation $q_{22}=0$ is:
\begin{eqnarray}
\label{fE9}
&&q_{22}=-\frac{\lambda_1J}{2} g_{22}+\lambda_3(G_{2;2;s}G^s+G^m_{;m}G_{2;2})=0\\
&&or\quad -\frac{\lambda_1 J}{2}-\frac{\lambda_3x^4}{\sqrt{\hat g}}{(\frac{\hat G_1}{\sqrt{\hat g}})}'=0
\end{eqnarray}

\newpage
{\it \underline{Appendix F:}}
\vskip 0.5em

In this appendix we derive an expression for $\bar q$ defined as $\bar q=q^{kl}G_kG_l$.

First, let us derive the expression for the invariant $I:=G_{k;l}G_{k;s}G^lG^s$ and $L_{12}=G_{k;l}G^{k;l}-G_{k;l}G^{l;k}$.
The invariant $I$ is the module of vector $I_k$ defined as $I_k=G_{k;s}G^s$, which has only two components -
 $I_0$ and $I_1$.
\begin{eqnarray}
\label{fF2}
&&I=(I_0)^2g^{00}+(I_1)^2g^{11}
\end{eqnarray}

\begin{eqnarray}
\label{fF3}
&&I_0=G_{0;s}G^s=G_{0;0}G_0g^{00}+G_{0;1}G_1g^{11}\nonumber\\
&&=\underline{-\Gamma^1_{00}G_1G_0g^{00}}+(G_{0,1}-\underline{\Gamma^0_{01}G_0})G_1g^{11}=G_{0,1}G_1g^{11}
\end{eqnarray}
In the expression above, the underlined terms cancel each other.
For the radial component $I_1$ we get:
\begin{eqnarray}
\label{fF4}
&&I_1=G_{1;s}G^s=G_{1;0}G_0g^{00}+G_{1;1}G_1g^{11}\nonumber\\
&&=-\Gamma^0_{01}G_0^2g^{00}+(G_{1,1}-\Gamma^1_{11}G_1)G_1g^{11}=\nonumber\\
&&-\frac{g_{0,1}}{2g_0}G_0^2-\frac{1}{2}{(\frac{G_1^2}{g_1})}_{,1}=
-\frac{1}{2}[\frac{g_{0,1}}{g_0}G_0^2+{(\frac{G_0^2}{g_0})}_{,1}]\nonumber\\
&&=-\frac{ {(G_0^2)}_{,1}}{2g_0}=
-\frac{G_0G_{0,1}}{g_0}
\end{eqnarray}
Substituting eq (\ref{fF3}) and (\ref{fF4}) in (\ref{fF2}) we get the final expression for I:
\begin{eqnarray}
\label{fF5}
&&I=(G_{0,1}G_1g^{11})^2g^{00}+(\frac{ {(G_0^2)}_{,1}}{2g_0})^2g_{11}=\nonumber\\
&&\frac{(G_{0,1})^2}{g_og_1}[\frac{G_1^2}{g_1}-
\frac{G_0^2}{g_0}]=-\frac{(G_{0,1})^2}{g_og_1}
\end{eqnarray}

We now derive the expression for $L_{12}:=L_1-L_2$:
\begin{eqnarray}
\label{fF6}
&&L_{12}=G_{k;l}G^{k;l}-G_{k;l}G^{l;k}\nonumber\\
&&=[(G_{0;0})^2(g^{00})^2+(G_{1;1})^2(g^{11})^2+(G_{0;1})^2g^{00}g^{11}+
(G_{1;0})^2g^{00}g^{11}]\nonumber\\
&&-[(G_{0;0})^2(g^{00})^2+(G_{1;1})^2(g^{11})^2+2G_{0;1}G_{1;0}g^{00}g^{11}]\nonumber\\
&&=G_{0;1})^2g^{00}g^{11}+(G_{1;0})^2g^{00}g^{11}+2G_{0;1}G_{1;0}g^{00}g^{11}\nonumber\\
&&=-\frac{1}{g_0g_1}(G_{0;1}-G_{1;0})^2=-\frac{(G_{0,1})^2}{g_og_1}
\end{eqnarray}
From which it follows that $I=L_{12}$.

Now we can derive the expression for $\bar q$.
We start with the equations obtained by the variation of the Lagrangian with respect to $g_{kl}$ eq. (\ref{f25}) derived in
Appendices A - eq. (\ref{fA20}), B - eq. (\ref{fB20}), and C - eq. (\ref{fC8a}). 

\begin{eqnarray}
\label{fF8}
&&G_kG_lq^{kl}:=G_kG_l\frac{1}{\sqrt{g}}\delta (\sqrt{g}L_G)/ \delta{g_{kl}}+T^{kl}G_kG_l=0\quad,\,\,where\nonumber\\
&&G_kG_lq_{kl}=G_kG_l\{\lambda_1[\frac7{18}(G^{m;n}G_{m;n})g_{kl}-{G^s}_{;k}G_{s;l}-\frac{5}{9}{G_k}^{;s}G_{l;s}]+\nonumber\\
&&\quad \,\,+\lambda_2[\frac7{18}(G^{m;n}G_{n;m})g_{kl}-\frac79 G_{k;s}{G^s}_{;l}-\frac79 G_{l;s}{G^s}_{;k}]\nonumber\\
&&\quad \,\, +\lambda_3[\frac7{18}({(G^m{;m})}^2g_{kl}-\frac79 G^s_s G_{k;l}-\frac79 G^s_s G_{l;k}]\nonumber\\
&&\quad \,\,+\lambda_1[\frac12{(G_{s;k}G_l+G_{s;l}G_k)}^{;s}+\frac5{18}{(G_{k;s}G_l+G_{l;s}G_k)}^{;s}\nonumber\\
&&\quad \,\, -\frac12 {(G_{k;l}G_s+G_{l;k}G_s)}^{;s}]\nonumber\\
&&\quad \,\,+\lambda_2[\frac19{(G_{m;n}G^n)}^{;m}g_{kl}+\frac5{18}{(G_{s;k}G_l+G_{s;l}G_k)}^{;s}\nonumber\\
&&\quad \,\,+\frac12{(G_{k;s}G_l+G_{l;s}G_k)}^{;s}-\frac12 {(G_{k;l}G_s+G_{l;k}G_s)}^{;s}]\nonumber\\
&&\quad \,\,+\lambda_3[-\frac89{(G^m_{;m}G^s)}_{;s}g_{kl}+\frac79{(G^m_{;m}G_k)}_{;l}+\frac79{(G^m_{;m}G_l)}_{;k}]\nonumber\\
&&\quad \,\,+\frac19T(g_{kl}+5G_kG_l)]\}
\end{eqnarray}

Due to the requirement $G_kG^k=1$, all the folowing invariants are zero:
\begin{eqnarray}
\label{fF9}
&&G_{k;l}G^kG^l=0;\quad G_{k;m}{G^l}_{;m}G^kG^l=0;\quad G_{k;m}{G^m}_{;l}G^kG^l=0 
\end{eqnarray}

Opening the parentheses we will get this expression:
\begin{eqnarray}
\label{fF10}
&&\bar q:=q_{ij}G^iG^j=\lambda_1[\frac{7}{18}L_1-J_1]+\lambda_2\frac{7}{18}L_2+\lambda_3\frac{7}{18} L_3\nonumber\\
&&+\lambda_1[{G^s}_{;k;s}G^k+\frac{10}{18} {G_{k;s}}^{;s} G^k-G_{k;l;s}G^kG^lG^s]\nonumber\\
&&+\lambda_2[\frac{J}{9}+\frac{5}{9}{G^s}_{;k;s}G^k+{G_{k;s}}^{;s}G^k-G_{k;l;s}G^kG^lG^s]\nonumber\\
&&\lambda_3[-\frac{8\bar J}{9}+\frac{14}{9}G^m_{;m;k}G^k ] + \frac{6T}{9}
\end{eqnarray}

We now write the following identities:
\begin{eqnarray}
\label{fF11}
&&{G^s}_{;k;s}G^k=({G^s}_{;k}G^k)}_{;s}-{G_{s;k}G^{k;s}=J-L_2\nonumber\\
&&{G_{k;s}}^{;s}G^k=(G_{k;s}G^k)^{;s}-G_{s;k}G^{s;k}=-L_1\nonumber\\
&&G^m_{;m;k}G^k={(G^m_{;m}G^k)}_{;k}-{(G^m_{;m})}^2=\bar J-L_3\nonumber\\
&&G_{k;l;s}G^kG^lG^s=[{(G_{k;l}G^k)}_{;s}G^lG^s-G_{k;l}G^k_{;s}]G^lG^s\nonumber\\
&&=-G_{k;l}G^k_{;s}G^lG^s=-L_{12} 
\end{eqnarray}
The derivation of the last identity was given in the beginning of this appendix.

Using these formulas we can rewrite the expression for $\bar q$, eq. (\ref{fF10}) as following:
\begin{eqnarray}
\label{fF12}
&&\bar q:=q_{ij}G^iG^j\nonumber\\
&&\quad =\frac{7}{18}L-\lambda_1(L_1-L_2)+\lambda_1(J-L_2-\frac{5}{9}L_1+L_1-L_2)\nonumber\\
&&+\lambda_2(\frac{6}{9}J-\frac{5}{9}L_2-L_1+L_1-L_2)+\lambda_3(-\frac{8}{9}\bar J+\frac{14}{9}\bar J-\frac{14}{9}L_3)\nonumber\\
&&+\frac{2}{3}T
\end{eqnarray}
Or, substituting $T=L-\lambda_2J+\lambda_3\bar J$ and using that
$\lambda_2=-\lambda_1-\lambda_3$ we get the final expression for $\bar q$:
\begin{eqnarray}
\label{fF13}
&&\bar q=q_{ij}G^iG^j=-\frac{L}{2}+\lambda_1(L_1-L_2)+\lambda_1J\nonumber\\
&& or\quad \bar q=\frac{\lambda_1}{2}L_{12}-\frac{\lambda_3}{2}L_{32}+\lambda_1J
\end{eqnarray}
which is the goal of this Appendix.

\newpage
{\underline {Appendix G}}
\vskip 1em

In this appendix we show that the solution obtained in the "Static sphere-symmetrical solution" section satisfies
 the equation $q_{00}=0$, eq. (\ref{f25}). Writing the equation $q_{00}=0$ in terms of the components we will get:
\begin{eqnarray}
\label{fG1}
&&q_{00}=0 \quad or \nonumber\\
&&\quad \lambda_1[\frac{7}{18}L_1g_{00}-G_{s;0}G^s_{;0}-\frac{5}{9}G_{0;s}{G_{0;}}^s]\nonumber\\
&&+\lambda_2[\frac{7}{18}L_2g_{00}-\frac{14}{9}G_{0;s}{G_{0;}}^s]\nonumber\\
&&+\lambda_3[\frac{7}{18}L_3-\frac{14}{9}G^m_{;m}G_{0;0}]\nonumber\\
&&+\lambda_1[(G^s_{;0;s}G_0+G_{0;s}G^s_{;0})+\frac{5}{9}({G_{0;s;}}^sG_0+G_{0;s}{G_0}^{;s})\nonumber\\
&&\quad \quad-(G_{0;0;s}G^s+G^m_{;m}G_{0;0})]\nonumber\\
&&+\lambda_2[\frac{1}{9}Jg_{00}+\frac{5}{9}(G^s_{;0;s}G_0+G_{0;s}{G_{0;}}^s)+({G_{0;s;}}^sG_0+G_{0;s}{G_{0;}}^s)\nonumber\\
&&\quad \quad-(G_{0;0;s}G^s+G^m_{;m}G_{0;0})]\nonumber\\
&&+\lambda_3[-\frac{8}{9}\bar J g_{00}+\frac{14}{9} G^m_{;m}G_{0;0}]+T/9(g_{00}+5{G_0}^2=0
\end{eqnarray}

Substituting $T=L-\lambda_2J-\lambda_3\bar J$, combining similar terms and taking into account that
 $\lambda_2=-\lambda_1-\lambda_3$ we get:

\begin{eqnarray}
\label{fG2}
&&q_{00}=0 \quad or \nonumber\\
&&\lambda_1[-\frac{1}{2}Jg_0+(2G_{0;s}G^s_{;0}-G_{0;s}{G_0}^{;s}-G_{s;0}G^s_{;0})\nonumber\\
&&\quad \quad+\frac{4}{9}J({G_{0;s;}}^sG_0-{G_{s;0;}}^sG_0)]\nonumber\\
&&+\lambda_3[\frac{5}{9}\{(J+\bar J)G_0^2-G^s_{;0;s}G_0\}+(G_{0;s}G^s_{;0}-G_{0;s}G_0^{;s}) \nonumber\\
&&\quad \quad +G_{0;0;s}G^s+G_{0;0}G^s_{;s}-G_{0;s;}^sG_0]=0
\end{eqnarray}

We now derive an expression for eq. (\ref{fG2}) through new variables - see eq. (\ref{f36}) and the coordinate $x=1/r$.

For the first square bracket the first term equals to:
\begin{eqnarray}
\label{fG3}
&&-\frac{1}{2}Jg_0=-\frac{K_Tx^4}{{2G_B}^4}g_0
\end{eqnarray}

The following three terms have this expression:
\begin{eqnarray}
\label{fG4}
&&2G_{0;s}G^s_{;0}-G_{0;s}{G_0}^{;s}-G_{s;0}G^s_{;0}\nonumber\\
&&=2({G_{0;0}}^2g^{0;0}+G_{0;1}G_{1;0}g^11)-{G_{0;0}}^2g^{00}-{G_{0;1}}^2g^{11}-{G_{0;0}}^2g^{00}-{G_{1;0}}^2g^{11}\nonumber\\
&&=(2G_{0;1}G_{10}-G_{0;1}-G_{1;0})g^{11}=-(G_{0;1}-G_{1;0})^2g^{11}\nonumber\\
&&=-(G_{0,1}-G_{1,0})^2g^{11}=(G_0\,')^2g_0\frac{x^4}{\hat g}\nonumber\\
&&or \quad \nonumber\\
&&2G_{0;s}G^s_{;0}-G_{0;s}{G_0}^{;s}-G_{s;0}G^s_{;0}=\frac{{K_T}^2g_0x^4}{{G_B}^4}
\end{eqnarray}

And the last two terms of in the first square bracket of (\ref{fG2}) cancel each other out:
\begin{eqnarray}
\label{fG5}
&&{G_{0;s;}}^sG_0-{G_{s;0;}}^sG_0=g_0(G^{0;s}-G^{s;0})_{;s}=g_0[(G^{0;s}-G^{s;0})\sqrt{g}]_{,s}\nonumber\\
&&=g_0G_0(G_{0,1}g^{00}g^{11}\sqrt{g})_{,1}=-g_0G_0 \frac{1}{\sqrt{\hat g}}(G_0\,'\frac{1}{\sqrt{\hat g}})\,'x^4=0
\end{eqnarray}

So the final result for the first square bracket of eq. (\ref{fG2}) is:
\begin{eqnarray}
\label{fG6}
&&\lambda_1[\,...]=\frac{\lambda_1K_Tx^4}{{2G_B}^4}g_0
\end{eqnarray}

In the second square bracket the first two terms cancel each other - for the detailed derivation see 
eq. (\ref{fG10}) and (\ref{fG12})-(\ref{fG13c}):

\begin{eqnarray}
\label{fG7}
&&(J+\bar J){G_0}^2-G^s_{;0;s}G_0\nonumber\\
&&=x^4[(1+\frac{\lambda_1}{2\lambda_3}) \frac{{K_T}^2x^4}{{G_B}^4}{G_0}^2-
\frac{1}{\hat g}{G_0}^2\frac{1}{2}g_0\,'']=0
\end{eqnarray}

The last five terms of the second bracket have these expressions as functions of x:

\begin{eqnarray}
\label{fG8}
&&G_{0;s}G^s_{\,;0}-G_{0;s}{G_0}^{;s}=\frac{x^4}{\hat g}[({G_0}\,')^2 g_0-\frac{1}{4}g_0\,'({G_0}^2)']\\
\label{fG9}
&&G_{0;0;s}G^s+G_{0;0}G^s_{;s}=x^4[\frac{1}{4\hat g}{G_0^2}\,'g_0\,'+\frac{1}{2\hat g}\hat Gg_0\,''x]\\
\label{fG10}
&&{G_{0;s;}}^sG_0=x^4\frac{G_0^2}{2\hat g}g_0\,''
\end{eqnarray}

For the detailed derivation see below eq. (\ref{fG11}), eq. (\ref{fG12})-(\ref{fG13c}) and eq. (\ref{fG14})-(\ref{fG14b})

Combining them all - eq. (\ref{fG8}), (\ref{fG9}), (\ref{fG10}) - together we will have for the second bracket:
\begin{eqnarray}
\label{fG10a}
&&\lambda_3[\,...]=\frac{\lambda_1K_Tx^4}{{2G_B}^4}g_0
\end{eqnarray}
 
Equations (\ref{fG6}) and (\ref{fG10a}) prove that equation $q_{00}=0$ is satisfied.

\begin{eqnarray}
&&\nonumber\\
&&*************\nonumber\\
&&\nonumber
\end{eqnarray}
We now will derive the result of eq. (\ref{fG8}), (\ref{fG9}), (\ref{fG10})

Indeed, for the eq. (\ref{fG8}) we have:
\begin{eqnarray}
\label{fG11}
&&G_{0;s}G^s_{\,;0}-G_{0;s}{G_0}^{;s}\nonumber\\
&&=G_{0;s}(G_{m;0}-G_{0;m})g^{sm}=G_{0;s}(G_{m;0}-G_{0;m})g^{sm}\nonumber\\
&&=G_{0;s}(G_{m,0}-G_{0,m})g^{sm}=G_{0;1}(G_{1,0}-G_{0,1})g^{11}\nonumber\\
&&=(G_{0,1}-\Gamma^0_{01} G_0)(-G_{0,1})(-\frac{1}{g_1})\nonumber\\
&&=[({G_0}\,')^2 g_0-\frac{1}{4}g_0\,'({G_0}^2)']\frac{1}{\hat g}
\end{eqnarray}

\begin{eqnarray}
\label{fG12}
&&*************
\end{eqnarray}

For the eq. (\ref{fG9}) we have:
\begin{eqnarray}
\label{fG12a}
&&G_{0;0;s}G^s+G_{0;0}G^s_{;s}\nonumber\\
&&=(-\Gamma^1_{00}G_{1;0}-\Gamma^1_{00}G_{0;1})G_0g^{00}+(G_{0;0},1-2\Gamma^0_{01}G_{0;0})G_1G^{11}\nonumber\\
&&\quad +(-\Gamma^1_{00}G_1)(G_{1,1}g^{11}-\Gamma^1_{00}g^{00}G_1-\Gamma^1_{11}g^{11}-2\Gamma^1_{22}g^{22})
\end{eqnarray}
In this case it is beneficial to switch to nornalized variables:
\begin{eqnarray}
&&\bar G_0=\frac{G_0}{\sqrt{g_0}} \quad and \quad  \bar G_1=\frac{G_1}{\sqrt{g_1}}
\end{eqnarray}
In the new variables the components of the tensor $G_{i;j}$ have this form:
\begin{eqnarray}
&&G_{0;0}=-\Gamma^1_{00}G_1=\frac{g_{0,1}}{2g_0} \frac{g_0}{\sqrt{g_1}}\bar G_1\\
&&G_{1;0}=-\Gamma^0_{01}G_0=-\frac{g_{0,1}}{2g_0}\bar G_0 \sqrt{g_0}\\
&&G_{0;1}=G_{0,1}-\Gamma^0_{01} G_0=G_{0,1}-\frac{g_{0,1}}{2g_0} G_0=\sqrt{g_0} \bar G_{0,1}\\
&&G_{1;1}=G_{1,1}-\Gamma^1_{11} G_1=G_{1,1}-\frac{g_{1,1}}{2g_1} G_1=\sqrt{g_1} \bar G_{1,1}\\
&&{G^m}_{;m}=(G_1g^{11}\sqrt{g})_{,1})\frac{1}{\sqrt{g}}=-(\bar G_1 g_2\sqrt{g_0})_{,1}\frac{1}{g_2\sqrt{\hat g}}
\end{eqnarray}

Using these expressions, the expression (\ref{fG12}) can be continued as:
\begin{eqnarray}
\label{fG13}
&&G_{0;0;s}G^s+G_{0;0}G^s_{;s}=\frac{g_0}{g_1}[\frac{1}{4}(\frac{g_{0,1}}{g_0})^2{G_0}^2
-\frac{1}{2}(\frac{g_{0,1}}{g_0}){\bar G_0}{\bar G_{0,1}}\nonumber\\
&&\quad -(-\frac{1}{2} \frac{g_{0,1}}{g_0}\bar G_1 \frac{1}{\sqrt{g_1}})_{,1}\bar G_1 \sqrt{g_1}\nonumber\\
&&\quad +\frac{1}{2} \frac{g_{0,1}}{g_0}\bar G_1(\bar G_{1,1}+\frac{1}{2} \frac{g_{0,1}}{g_0}\bar G_1+
\frac{g_{2,1}}{g_2}\bar G_1)]\\
\end{eqnarray}

The terms $\bar G_{1,1} \bar G_1$ can be replaced with $\bar G_{0,1} \bar G_0$, which comes from condition $G_iG^i=1$.
\begin{eqnarray}
\label{fG13a}
&&=\frac{g_0}{g_1} \{[\frac{1}{4} (\bar G_0^2)_{,1} \frac{g_{0,1}}{g_0}
+\frac{1}{4} (\frac{g_{0,1}}{g_0})^2\bar G_0^2]\nonumber\\
&&+\bar G_1^2[\frac{1}{2}(\frac{g_{0,1}}{g_0})_{,1}-\frac{1}{4}\frac{g_{0,1}}{g_0}\frac{g_{1,1}}{g_1}
+\frac{1}{4}(\frac{g_{0,1}}{g_0})^2+\frac{1}{2}\frac{g_{0,1}}{g_0}\frac{g_{2,1}}{g_2}]\}\nonumber\\
&&=\frac{g_0}{g_1} \{\frac{1}{4}\frac{g_{0,1}}{g_0} (\bar G_0^2g_0)_{,1} \frac{1}{g_0}\nonumber\\
&&+\bar G_1^2[\frac{1}{2g_2}(\frac{g_{0,1}}{g_0}g_2)_{,1}-\frac{1}{4}\frac{g_{0,1}}{g_0}\frac{g_{1,1}}{g_1}+
\frac{1}{4}(\frac{g_{0,1}}{g_0})^2]\}
\end{eqnarray}
Differentiating the first term in the square bracket to be $(g_{0,1}g_2)_{,1}/g_0-(g_{0,1}/g_0)^2$
the expressiton above can be written as:
\begin{eqnarray}
\label{fG13b}
&&=\frac{g_0}{g_1} \{\frac{1}{4}\frac{g_{0,1}}{g_0} (\bar G_0^2g_0)_{,1} \frac{1}{g_0}\nonumber\\
&&+\bar G_1^2[\frac{1}{2g_2}\frac{(g_{0,1}g_2)_{,1}}{g_0}-(\frac{1}{4}\frac{g_{0,1}}{g_0}\frac{g_{1,1}}{g_1}+
\frac{1}{4}(\frac{g_{0,1}}{g_0})^2]\}
\end{eqnarray}

We will now switch to the new variable $\hat G=G_1^2g_0r/g_1=\bar G_1^2g_0$r, $\hat g=g_0g_1$, 
x=1/r and replacing $g_2$ with $r^2$:
\begin{eqnarray}
\label{fG13c}
&&=x^4[\frac{{g_0}'(G_0^2)\,'}{4\hat g}+\frac{\hat Gx{g_0}''}{2\hat g}-\frac{\hat Gx{\hat g}'}{4{\hat g}^2}]=
x^4[\frac{{g_0}'(G_0^2)\,'}{4\hat g}+\frac{\hat Gx{g_0}''}{2\hat g}]
\end{eqnarray}
where we use the fact that $\hat g=constant$.

\vskip 1em
\begin{eqnarray}
\label{fG14}
&&*************
\end{eqnarray}

For the eq. (\ref{fG10}) we have:
\begin{eqnarray}
\label{fG14a}
&&G_0G_{s;0;m}g^{sm}=G_0(G_{s;0,m}g^{sm}-\Gamma^p_{ms}G_{p;0}g^{ms}-\Gamma^p_{m0}G_{s;p}g^{ms})\nonumber\\
&&=G_0[G_{1;0,1}g^{11}-(\Gamma^1_{00}g^{00}+\Gamma^1_{11}g^{11}+\Gamma^1_{22}g^{22})G_{1;0}\nonumber\\
&&\quad -\Gamma^1_{00}G_{0;1}-\Gamma^0_{01}G_{1;0}]g^{11}\nonumber\\
&&=G_0g^{11}[(-\Gamma^0_{10,1}G_0-\Gamma^0_{10}G_{0,1})\nonumber\\
&&\quad -(\Gamma^1_{00}g^{00}+\Gamma^1_{11}g^{11}+\Gamma^1_{22}g^{22})(-\Gamma^0_{10}G_0)\nonumber\\
&&\quad -\Gamma^1_{00}(G_{0,1}-\Gamma^0_{10}G_0)g^{00}-\Gamma^0_{10}(-\Gamma^0_{10}G_0)]
\end{eqnarray}
The terms with $G_{0,1}$ cancel each other out. 
Opening the parentheses and substituting for 
$\Gamma^1_{00},\,\Gamma^1_{11}, \Gamma^1_{22}$ the  expressions through $g_0$, $g_1$ and $g_2$ we will get:
\begin{eqnarray}
&&=\frac{G_0^2}{g_1}[\Gamma^0_{10,1}+\frac{g_{0,1}}{2g_0}\Gamma^0_{10}-\frac{g_{1,1}}{2g_1}\Gamma^0_{10}
+\frac{g_{2,1}}{g_2}\Gamma^0_{10}]\nonumber\\
&&=\frac{G_0^2}{g_1}(\frac{\Gamma^0_{10}\sqrt{g_0}g_2}{\sqrt{g_1}})_{,1}
\frac{\sqrt{g_1}}{\sqrt{g_0}g_2}=\frac{G_0^2}{2\sqrt{\hat g}r^2}(\frac{g_{0,1}r^2}{\sqrt{\hat g}})_{,1}\nonumber
\end{eqnarray}
where we use $\hat g=g_0g_1$ and $g_2=r^2$.

Switching to the new variable (x=1/r) and remembering that the parameter $\hat g=constant$ we get the final result:
\begin{eqnarray}
\label{fG14b}
&&G_0G_{s;0;m}g^{sm}=\frac{G_0^2x^4}{2\hat g}g_0\,''
\end{eqnarray}

\newpage
{\underline {Appendix H}}
\vskip 1em
In this appendix we will show that the solution obtained in the "Static, spherically symmetric solution" section satisfy the equation
$Q_0=0$, eq. (\ref{f23a}).

\begin{eqnarray}
\label{fH1}
&&Q_0=0 \quad where\nonumber\\
&&Q_0=\lambda_1({G_{0;s;}}^s-{G^s}_{;0;s})+\lambda_3({G^m}_{;m:0}-{G^s}_{;0;s})+TG_0
\end{eqnarray}

Let us consider the first backet:
\begin{eqnarray}
\label{fH2}
&&({G_{0;s;m}}-G_{m;0;s})g^{sm}=\\
&&(G_{0;s}-G_{s;0})_{,m} g^{sm}-\Gamma^p_{m0}(G_{p;s}-G_{s;p})g^ms-\Gamma^p_{ms}g^{sm}(G_{0;p}-G_{p;0})\nonumber
\end{eqnarray}

In each parentheses the semicolon sign (;) could be replaced with the simple partial derivative (,) due to the fact that
$\Gamma$s inside parentheses cancel each other out. 

In the first $\Gamma^p_{s0}$, the indices p and s can 
only have these two combinations: p=0, m=s=1 or p=1, m=s=1. 
In the second $\Gamma^p_{sm}g^{sm}$, the parameter p=1. Taking these into account we get:
\begin{eqnarray}
\label{fH3}
&&=(G_{0,1}-G_{1,0})_{,1} g^{11}-\Gamma^0_{10}(G_{0,1}-G_{1,0})g^{11}-\underline{\Gamma^1_{00}(G_{1,0}-G_{0,1})g^{00}}\nonumber\\
&& \quad -(\underline{\Gamma^1_{00}g^{00}}+\Gamma^1_{11}g^{11}+2\Gamma^1_{22}g^{22})(G_{0,1}-G_{1,0})\nonumber
\end{eqnarray}
The underlined terms cancel each other.  The partial derivative by $x_0$ (,0) vanishes for the static problem. And replacing
$\Gamma$s with their expressions through the metric tensor we get:
\begin{eqnarray}
\label{fH4}
&&=(-\frac{1}{g_1})[G_{0,1,1}-\frac{g_{0,1}}{2g_0}G_{0,1}-\frac{g_{1,1}}{2g_1}G_{0,1}
+\frac{g_{2,1}}{g_2}G_{0,1}\nonumber\\
&&=(-\frac{1}{g_1})(\frac{G_{0,1}g_2}{\sqrt{g_0g_1}})_{,1}\frac{\sqrt{g_0g_1}}{g_2}\nonumber
\end{eqnarray}
Switching to a new variable x=1/r and remembering that $g_2=r^2$ and that $\hat g:={g_0g_1}=constant$ we get:
\begin{eqnarray}
\label{fH5}
&&\lambda_1({G_{0;s;}}^s-{G^s}_{;0;s})=(-\frac{\lambda_1}{g_1})(G_0)\,''x^4
\end{eqnarray}
 
We can now consider the second term in eq. (\ref{fH1}). The first part ($G^m_{;m;0}$) vanishes due to the fact
that it is a partial derivative by $x_0$. The second part, using eq. (\ref{fG10}) - for its derivation see 
eq. (\ref{fG14})-(\ref{fG14b}) - can be written as:
\begin{eqnarray}
\label{fH6}
&&-\lambda_3G_{s;0;m}g^{sm}=-\frac{\lambda_3G_0x^4}{2\hat g}g_0\,''
\end{eqnarray}

Combining eq. (\ref{fH5}) and (\ref{fH6}) above,  and using the expression for the invariant T, eq. (\ref{f52})
the equation $Q_0=0$ can be written as:
\begin{eqnarray}
\label{fH7}
&&\lambda_1({G_{0;s;}}^s-{G^s}_{;0;s})+\lambda_3({G^m}_{;m:0}-{G^s}_{;0;s})+TG_0\nonumber\\
&&or \quad=(-\frac{x^4\lambda_1}{g_1})(G_0)\,''-\frac{\lambda_3G_0x^4}{2\hat g}g_0\,''+\frac{\lambda_3x^4}{\hat g}
(1+\frac{\lambda_1}{2\lambda_3})G_0=0\nonumber
\end{eqnarray}
The solution for $G_0$ is linear of x and thus the first term equals zero. 
Using expression for $g_0$, eq. (\ref{f50}), as a quadratic function of variable "x",
the second and the third terms cancel each other, which proves that equation $Q_0=0$ is satisfied.

\end{document}